\documentclass[runningheads]{llncs}
\usepackage[T1]{fontenc}
\usepackage{graphicx}
\usepackage{amsmath}
\usepackage[caption=false]{subfig}
\usepackage[table]{xcolor}
\usepackage{colortbl}
\usepackage{svg}
\usepackage{threeparttable}
\definecolor{groupbg}{RGB}{200,220,240}
\definecolor{rowbg}{HTML}{F7F9FC}
\usepackage{enumitem}
\usepackage{adjustbox}
\usepackage{sidecap}
\usepackage{booktabs}
\usepackage{amssymb}
\usepackage{hyperref}

\makeatletter
\renewcommand{\section}{\@startsection{section}{1}{\z@}%
  {-1ex \@plus -0.5ex \@minus -0.2ex}
  {1.0ex \@plus 0.2ex}
  {\normalfont\Large\bfseries}} 
\makeatother
\AtBeginDocument{%
  \setlength{\abovedisplayskip}{4pt}
  \setlength{\belowdisplayskip}{4pt}
  \setlength{\abovedisplayshortskip}{2pt}
  \setlength{\belowdisplayshortskip}{2pt}
  \setlength{\jot}{3pt}
}
\makeatletter
\renewcommand\@openbib@code{\setlength{\itemsep}{0pt}\setlength{\parskip}{0pt}}
\makeatother

\usepackage{todonotes}

\begin{document}
\title{Modeling and Chasing the Energy-Efficiency Sweet Spots in Modern GPUs
}
\author{Ayesha Afzal\inst{1}\orcidID{0000-0001-5061-0438} \and
Markus Manfred Li\inst{2}\orcidID{0009-0004-0908-3460} \and
Michael Panzlaff\inst{1}\orcidID{0009-0009-2993-7001} 
}
\authorrunning{A. Afzal et al.}
%
\institute{Erlangen National High Performance Computing Center (NHR@FAU) \and
Department of Computer Science, FAU 
Erlangen-Nürnberg, Germany
}
\maketitle             

\begin{abstract}
Energy consumption is a key limitation in high-performance computing on heterogeneous CPU--GPU systems. This work studies how hardware configuration affects energy-to-solution under realistic workloads.
We study energy efficiency regimes using molecular dynamics benchmarks (GROMACS and AMBER) and a stress-test benchmark (FIRESTARTER) on systems with A40, A100, H100, and H200 GPUs and Intel Ice Lake CPU, varying frequency scaling and power cap. 
We show that energy-to-solution exhibits workload- and architecture-dependent transitions between efficient and inefficient regimes, driven by nonlinear GPU power-frequency scaling.
We introduce an interpretable analytical model that decomposes GPU power into linear and nonlinear components, identifying a workload- and architecture-dependent transition frequency beyond which efficiency degrades. The model fits empirical data with low error and highlights the role of baseline power, nonlinear power behavior, and transition frequency as the dominant parameters governing energy efficiency.
Power capping is generally less effective for efficiency tuning than frequency reduction, especially for workloads that operate far from thermal design power. 
Overall, energy-efficient HPC execution is a configuration-dependent problem with identifiable regime shifts, and we provide model-driven guidance for selecting operating points.
\end{abstract}


\section{Introduction and related work}
Energy efficiency has emerged as a primary limiting factor in high-performance computing (HPC), increasingly constraining system scalability under fixed power budgets rather than peak computational capability. This challenge is particularly pronounced in heterogeneous CPU--GPU systems, where multiple hardware control parameters jointly determine performance and energy to solution.
    A central difficulty arises from the high-dimensional configuration space exposed by modern HPC systems, including GPU dynamic voltage and frequency scaling (DVFS)~\cite{Krzywaniak:2020}, power caps~\cite{Han:2025}, and empirical power modeling~\cite{Horowitz:2005,Weste:2015,Li:2009,Leng:2013}. These strategies exhibit nonlinear GPU power scaling due to voltage--frequency coupling and workload dependence, yet most studies rely on synthetic kernels, limiting their applicability to real-world applications. Few existing approaches implicitly assume smooth and monotonic relationships between performance, power, and energy efficiency. However, such assumptions frequently break down for scientific workloads with heterogeneous compute and memory behavior.
Empirical or machine-learning-based energy models exist but often lack interpretability and hardware-level insight, creating a gap between predictive accuracy and physical understanding for realistic HPC workloads.
In this work, we focus on molecular dynamics (MD) applications, represented by GROMACS~\cite{Abraham:2015,AfzalHW:2025:2} and AMBER~\cite{Case:2005}, complemented by the synthetic stress-test benchmark FIRESTARTER~\cite{Hackenberg:2013}. These workloads are executed on multiple GPU architectures (A40, A100, H100, H200) and an Intel Ice Lake CPU platform. Collectively, they span compute-bound, memory-bound, and thermally constrained regimes, enabling a unified analysis of energy behavior across diverse operating conditions.

\textit{\textbf{Contributions}}\quad
This paper makes the following contributions:
\begin{itemize}[topsep=0pt]
    \item We systematically evaluate energy-to-solution using molecular dynamics workloads (GROMACS, AMBER) and a synthetic stress-test (FIRESTARTER), covering GPU frequency scaling and power cap configurations.
    \item We introduce an analytic GPU power–frequency model fitted to real application data with low error across GPU architectures (A40, A100, H100, H200), decomposing power into baseline, linear, and nonlinear terms; it identifies workload- and architecture-dependent transition frequencies beyond which efficiency collapses due to quadratic power growth, while CPUs (Intel Ice Lake) exhibit near-linear scaling.
    \item We integrate empirical analysis with analytic modeling to derive predictive operating regimes for energy-efficient HPC, showing that energy increases nonlinearly beyond workload-dependent transition points despite diminishing performance gains. 
    \item We show that frequency scaling consistently outperforms power capping for energy optimization, particularly for compute-light workloads, and show that optimal operating points require workload-aware hardware configuration.
\end{itemize}
Our results establish energy-efficient HPC execution as a configuration-dependent, regime-based problem, providing a principled foundation and model-driven guidance for selecting operating points for MD and stress-test workloads.

\textit{\textbf{Overview}}\quad
The remainder of this paper is organized as follows: In Sec.~\ref{sec:setup}, we describe the benchmark workloads, hardware testbed, software configuration, and experimental methodology used in our study. Sec.~\ref{sec:powermodel} presents the analytical GPU and CPU power-frequency model. Sec.~\ref{sec:eval} presents the evaluation of energy-to-solution across architectures, comparing frequency tuning, and power cap strategies. Sec.~\ref{sec:conclusion} summarizes the main findings and outlines future work.

\section{Workloads, testbed and experimental setup}\label{sec:setup}
This section describes the experimental platforms, workloads, measurement methodology, and parameter exploration used to analyze energy behavior in heterogeneous CPU--GPU systems.

\begin{table*}[t]
\caption{Selected MD benchmarks across a range of biomolecular systems and sizes.}
\label{tab:mdbench}
\centering
\renewcommand{\arraystretch}{1.2}
\setlength{\tabcolsep}{6pt}
\rowcolors{2}{rowbg}{white}

\begin{threeparttable}
\scriptsize

\begin{tabular}{|
>{\centering\arraybackslash}p{2.4cm}|
>{\centering\arraybackslash}p{1.1cm}|
>{\centering\arraybackslash}p{1.6cm}|
>{\centering\arraybackslash}p{1.1cm}|
>{\centering\arraybackslash}p{1.75cm}|
>{\centering\arraybackslash}p{1cm}|
}

\hline

\rowcolor{groupbg}
\multicolumn{2}{|c|}{\textbf{GROMACS GPU}} &
\multicolumn{2}{c|}{\textbf{AMBER GPU}} &
\multicolumn{2}{c|}{\textbf{AMBER CPU}} \\
\hline

\rowcolor{groupbg}
\textbf{Benchmark} &
\textbf{Atoms} &
\textbf{Benchmark} &
\textbf{Atoms} &
\textbf{Benchmark} &
\textbf{Atoms} \\
\hline
2md\_start0  & 20,248 & TRPCage & 304 & Thioredoxin & 14,093 \\
rnanvt  & 31,889 & Myoglobin & 2,492 & JAC & 23,558 \\
PI\_large\_test & 80,289 & JAC$\star$ & 23,558 & JAC (pmemd) & 23,558 \\
FL\_md1\_berendsen  & 170,320 & Nucleosome & 25,095 & DHFR & 23,930 \\
eag1  & 615,924 & Factor IX$\star$ & 90,906 & Factor IX & 90,906 \\
stmv\_pme\_nvt & 1,066,628 & Cellulose$\star$ & 408,609 &  &  \\
 &  & STMV$\star$ & 1,067,095 &  &  \\
\hline

\end{tabular}
\begin{tablenotes}[flushleft]\footnotesize
\item[$\star$] GPU runs use standard NPT or NVE ensembles following typical MD workflows.
\end{tablenotes}
\end{threeparttable}
\end{table*}
\textit{\textbf{Workloads}}\quad
We evaluate two workload classes: (1) \textit{molecular dynamics (MD) application workloads GROMACS and AMBER} and (2) the \textit{synthetic stress-test benchmark FIRESTARTER}. 
The MD benchmark suite spans workloads from small, GPU-underutilized benchmarks to large, throughput- and memory-bound simulations, differing in parallelization strategies and compute intensities. This enables analysis of heterogeneous performance and energy behavior across architectures.  
Detailed configurations and atom counts are summarized in Table~\ref{tab:mdbench}.
We evaluate six publicly available \textbf{GROMACS}\footnote{\url{https://manual.gromacs.org/documentation/current}} benchmarks including solvent boxes, membrane proteins, and viral benchmarks. Larger benchmarks are increasingly sensitive to memory bandwidth. Inputs are available in the Artifact Description\footref{foot:AD}, and simulations follow benchmark settings.
We evaluate eleven GPU and five CPU standard \textbf{AMBER}\footnote{\url{https://ambermd.org/GPUPerformance.php}} benchmarks spanning small peptides to large biomolecular assemblies. This span enables the comprehensive CPU--GPU energy analysis from latency-dominated workloads (e.g., smaller TRPCage and Myoglobin benchmarks, where kernel launch overhead and per-step latency dominate) to throughput- and memory-bound regimes (e.g., larger Cellulose and STMV benchmarks, where long-range electrostatics and memory bandwidth become limiting).
The CPU benchmarks represent standard MPI/OpenMP workloads that are predominantly compute-bound on Ice Lake CPU architectures, and primarily probing parallel efficiency and cache/memory hierarchy behavior.
To study worst-case sustained power behavior, we additionally use \sloppy\textbf{FIRESTARTER}\footnote{\url{https://tu-dresden.de/zih/firestarter}}, a synthetic stress benchmark based on architecture-specific low-level kernels designed to maximize utilization of compute units, caches, and memory subsystems. Unlike application workloads, it targets near-TDP operation and is commonly used for infrastructure validation and energy-efficiency studies. 

\begin{table*}[t]
\caption{Key specifications of NVIDIA GPUs used in the experiments. }
\centering
\renewcommand{\arraystretch}{1.2}
\setlength{\tabcolsep}{8pt}
\rowcolors{2}{rowbg}{white}
\begin{threeparttable}
\scriptsize
\begin{tabular}{|
>{\centering\arraybackslash}p{0.58cm}| 
>{\centering\arraybackslash}p{0.7cm}| 
>{\centering\arraybackslash}p{1.45cm}| 
>{\centering\arraybackslash}p{0.75cm}| 
>{\centering\arraybackslash}p{1.28cm}| 
>{\centering\arraybackslash}p{0.65cm}| 
>{\centering\arraybackslash}p{0.72cm}| 
>{\centering\arraybackslash}p{0.7cm}| 
}

\hline

\rowcolor{groupbg}
\textbf{GPUs} &
\textbf{MEM clock std.} &
\textbf{Graphics clock min--max} &
\textbf{MEM bandwidth} &
\textbf{Power cap min--max} &
\textbf{TDP} &
\textbf{MEM type} &
\textbf{MEM size} \\

\rowcolor{groupbg}
&
\makebox[0pt][c]{[GHz]} &
\makebox[0pt][c]{[GHz]} &
\makebox[0pt][c]{[GB/s]} &
\makebox[0pt][c]{[W]} &
\makebox[0pt][c]{[W]} &
&
\makebox[0pt][c]{[GB]} \\

\hline
A40  & 7.251 & 0.21 / 1.74  & 798   & 100--300 & 300 & GDDR6 & 48 \\
A100 & 1.215 & 0.21 / 1.41  & 1,555 & 100--400 & 400 & HBM2  & 40 \\
H100 & 1.593 & 0.345 / 1.98 & 3,352 & 350--700 & 700 $\dagger$ & HBM3  & 80 \\
H200 & 1.41  & 0.40 / 1.98  & 4,800 & 400--700 & 700 $\dagger$ & HBM3e & 141 \\
\hline
\end{tabular}

\begin{tablenotes}[flushleft]\footnotesize
\item[$\dagger$] \emph{Thermal Design Power (TDP)} denotes the typical full-load power draw without enforced power limits. Default power cap on the Helma\footref{foot:helma} configuration is 500~W.
\end{tablenotes}

\end{threeparttable}
\label{tab:specs}
\end{table*}
\textit{\textbf{HPC platforms and architectures}}\quad
Experiments were conducted on heterogeneous HPC systems with Intel Ice Lake CPUs and NVIDIA GPUs (A40, A100, H100, and H200); key specifications are summarized in Table~\ref{tab:specs}.
Three clusters were at our disposal:
\begin{enumerate}[topsep=0pt]
    \item Helma\footnote{\url{https://doc.nhr.fau.de/clusters/helma}\label{foot:helma}} comprising eight NVIDIA H100 or H200 GPUs per node
    \item Alex\footnote{\url{https://doc.nhr.fau.de/clusters/alex}} comprising eight NVIDIA A40 or A100 GPUs per node
    \item Fritz\footnote{\url{https://doc.nhr.fau.de/clusters/fritz}} comprising dual-socket Intel Xeon Platinum 8360Y (Ice Lake) nodes with 72 cores per node, base frequency 2.4~GHz (up to 3.5~GHz turbo), 256~GB DDR4-3200 memory, AVX-512 vector units, and SMT disabled.
\end{enumerate}
The systems span multiple architectural generations and differ in compute throughput, memory bandwidth, thermal characteristics, and power-management behavior, allowing us to assess whether energy-efficiency trends generalize across platforms. Even for GPU-accelerated MD codes, CPU-side configuration and CPU--GPU interaction significantly influence performance and energy efficiency (more so for GROMACS than AMBER)~\cite{Afzal:2023:2}.

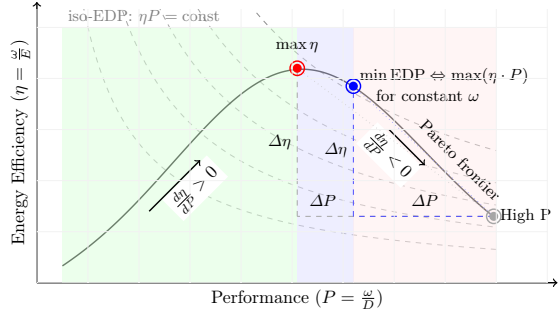
\begin{SCfigure}[][!t]
\centering
\begin{adjustbox}{max width=0.6\linewidth}
\begin{tikzpicture}[scale=1]

\draw[->, thick] (0,0) -- (10.2,0)
node[midway, below] {Performance ($P = \frac{\omega}{D}$)};

\draw[->, thick] (0,0) -- (0,5.5)
node[midway, above, rotate=90] {Energy Efficiency ($\eta = \frac{\omega}{E}$)};

\clip (0,0) rectangle (10.2,5.35);

\foreach \c in {6,10,14,18} {
    \draw[gray!70, dashed, smooth, domain=0.5:9, samples=100]
    plot (\x, {\c/(\x+0.2)});
}

\node[gray!100] at (2.1,5.2) {\small iso-EDP: $\eta P = \mathrm{const}$};
\draw[gray!90, dashed, smooth, domain=0.5:9, samples=100]
plot (\x, {23.8/(\x+0.2)});

\draw[thick, black, smooth, domain=0.5:9, samples=200]
plot (\x, {5*(\x/(1+\x)) * exp(-0.08*(\x-5)^2)});


\fill[green!15, opacity=0.5]
(0.5,0) -- (0.5,5) -- (5.1,5) -- (5.1,0) -- cycle;
\node[rotate=45, fill=white, opacity=1, text opacity=1, inner sep=2pt] at (3.1,1.85) {$\frac{d\eta}{dP} > 0$};

\fill[blue!13, opacity=0.5]
(5.1,0) -- (5.1,5) -- (6.2,5) -- (6.2,0) -- cycle;

\fill[red!8, opacity=0.5]
(6.2,0) -- (6.2,5) -- (9,5) -- (9,0) -- cycle;
\node[rotate=-45, fill=white, opacity=1, text opacity=1, inner sep=2pt] at (6.9,2.5) {$\frac{d\eta}{dP} < 0$};

\node[rotate=-45] at (8.3,2.4) {\small Pareto frontier};


\draw[dashed, gray!70] (8.95,1.3) -- (5.1,1.3);   
\draw[dashed, gray!70] (5.1,1.3) -- (5.1,4.2);   

\node[below] at (5.6,1.85) {\small $\Delta P$};
\node[left]  at (5.1,2.75) {\small $\Delta \eta$};

\draw[gray!40, dotted] (8.95,1.3) -- (5.1,4.2);

\draw[dashed, blue!70] (8.95,1.3) -- (6.2,1.3);   
\draw[dashed, blue!70] (6.2,1.3) -- (6.2,3.85);   

\node[below] at (7.6,1.85) {\small $\Delta P$};
\node[left]  at (6.2,2.6) {\small $\Delta \eta$};

\draw[blue!40, dotted] (8.95,1.3) -- (6.2,3.85);


\fill[red] (5.1,4.2) circle (2.5pt);
\draw[red, thick] (5.1,4.2) circle (4pt);
\node[above] at (5.1,4.4) {\small $\max \eta$};

\fill[blue] (6.2,3.85) circle (2.5pt);
\draw[blue, thick] (6.2,3.85) circle (4pt);
\node[right] at (6.15,3.85) {
\begin{tabular}{l}
\small $\min \mathrm{EDP} \Leftrightarrow \max(\eta \cdot P)$ \\
\small \quad $\mathrm{for~constant~}\omega$
\end{tabular}
};

\fill[gray!70] (8.95,1.3) circle (2.5pt);
\draw[gray!70, thick] (8.95,1.3) circle (4pt);
\node[right] at (8.95,1.3)
{\small High P};

\draw[->, thick] (2.2,1.4) -- (3.2,2.4);
\draw[->, thick] (6.9,3) -- (7.6,2.3);

\node at (8.5,0.6) {\small };
\node at (1.5,0.6) {\small };


\draw[gray!10] (0,0) grid (10,6);

\end{tikzpicture}
\end{adjustbox}
\caption{Performance–energy efficiency trade-off, illustrating iso-EDP contours and the Pareto frontier (lying between the red and grey points). The minimum-EDP (blue point) corresponds to the point of tangency between an iso-EDP contour and the Pareto frontier.}
\label{fig:pareto-edp}
\end{SCfigure}

\textit{\textbf{Metrics}}\quad
We analyze four metrics: (1) \textit{Performance} in ns/day\footnote{Performance is normalized by atom count for cross-benchmark comparability.} (physical time simulated per wall-clock day) for MD workloads and FLOPS (floating-Point Operations per Second) for FIRESTARTER, (2) \textit{average power consumption} in W, (3) total \textit{Energy-to-solution} in J, and (4) \textit{Energy-delay product (EDP)} in $ \mathrm{J \cdot s} $ for a fixed workload. We use EDP, though other formulations, such as ED$^2$P, are also valid.
Performance is defined as work $\omega$ normalized by delay $D$ (runtime), and energy efficiency as useful work per energy $E$. 
Configurations are analyzed using Pareto-optimal performance--energy trade-offs (Fig.~\ref{fig:pareto-edp}). We distinguish three operating regimes:
(1) Energy efficiency-improving (green): higher performance increases efficiency;
(2) Balanced (blue): the Pareto-optimal knee region;
and (3) Energy-inefficient (red): further performance gains incur disproportionate energy use.
Optimal operating points are identified using (1) \textit{\textbf{maximum energy efficiency $\eta$}} (minimum Energy-to-solution; red point), and (2) \textbf{minimum EDP} (energy–performance trade-offs; blue point. It enables systematic identification of workload-dependent optimal operating points and transition boundaries across architectures.

\textit{\textbf{Configuration parameters sweeps}}\quad
We systematically vary CPU and GPU control parameters to study their impact on energy-to-solution.
(1) \textbf{Frequency scaling:} GPU graphics and CPU core frequencies are varied across their supported ranges using \texttt{nvidia-smi --\kern0pt-lock-gpu-clocks} and \sloppy\texttt{--\kern0pt-cpu-freq:performance}, respectively. GPU memory clocks remain at their default operational values and not reduced to idle-state (P8) levels, as the latter would result in impractically low memory bandwidth. Clocks are fixed during execution and no explicit power cap is applied.
(2) \textbf{Power cap:} GPU power limits are controlled using \texttt{nvidia-smi --\kern0pt-power-limit}, across the supported range while retaining default clock frequencies. In contrast to frequency scaling, power caps impose an upper power bound to help with thermal constraints or datacenter power budgeting and trigger throttling only when limits are reached. Values below  \emph{minimum cap} are invalid. 

\textit{\textbf{Software configuration}}\quad
All experiments used reproducible HPC software stacks summarized in Table~\ref{tab:software}. GROMACS 2024.4 was executed in a hybrid MPI+OpenMP configuration (one MPI rank per GPU, 16 OpenMP threads) with full GPU offloading for all major MD kernels using binary \texttt{.tpr} inputs (preprocessed with \texttt{grompp}). AMBER24 GPU runs used \texttt{pmemd.cuda}, while AMBER20 CPU baselines used MPI-enabled \texttt{pmemd} and \texttt{sander}. FIRESTARTER executed a standalone CUDA stress kernel with fixed computational intensity to generate deterministic GPU load under controlled conditions.

\textit{\textbf{Experimental methodology}}\quad
All codes were compiled with architecture-specific optimization flags targeting the native capabilities of the respective GPUs.
Experiments were executed sequentially across frequency settings and benchmarks on reserved SLURM nodes with dedicated GPU allocation per job (\texttt{CUDA\_VISIBLE\_DEVICES=0}) to avoid resource contention and scheduling interference. GPU power and utilization were sampled via \texttt{nvidia-smi} at 100\,ms intervals, while CPU power was measured using \texttt{likwid-perfctr -g ENERGY}. Thread and core affinity were enforced for all MPI/OpenMP workloads to ensure deterministic CPU scheduling and for reproducibility. Each configuration was repeated three times and evaluated under stable runtime conditions. Power measurements were averaged only during steady-state utilization phases (85\% GPU utilization for MD workloads and 100\% for FIRESTARTER) to avoid startup and transient effects. Observed variability remained below 5\% in most cases, with slightly higher fluctuations in selected FIRESTARTER runs on Helma due to presumably thermal conditions, hardware variability, and architecture-specific DVFS behavior. 
Our analysis focuses on stable and reproducible behavioral trends governing energy efficiency across workloads and hardware generations, rather than recovering exact optimal frequencies at single-watt precision.
Execution timestamps and application runtime logs were recorded for all configurations. 
GPU-only and CPU-only measurements reflect accelerator and on-chip power consumption and exclude system baseline power~\cite{AfzalHW:2026,AfzalHW:2025:1}. During GPU-only execution on FIRESTARTER, CPU contributions were negligible, and GPU utilization was 100\%. 
To capture system-level effects, IPMI telemetry was sampled for node-level power, thermal, and cooling measurements. On Alex under mixed workloads, approximately 27\% of total node power was attributed to non-compute subsystems, including ~6\% for fans and the remainder to network components, storage, memory, motherboard subsystems, and peripherals. On newer architectures like H200, memory can significantly contribute to total power.


\begin{table*}[t]
\caption{Software stacks, execution models, and workload configurations.}
\label{tab:software}
\centering
\renewcommand{\arraystretch}{1.2}
\setlength{\tabcolsep}{6pt}
\rowcolors{2}{rowbg}{white}

\begin{threeparttable}
\scriptsize

\begin{tabular}{|
>{\centering\arraybackslash}p{1.4cm}|
>{\centering\arraybackslash}p{2.8cm}|
>{\centering\arraybackslash}p{3.7cm}|
>{\centering\arraybackslash}p{2.9cm}|
}

\hline

\rowcolor{groupbg}
\textbf{Component} &
\textbf{GROMACS} &
\textbf{AMBER} &
\textbf{FIRESTARTER} \\
\hline

\textbf{Execution model} &
Single-node GPU; 1 MPI rank/GPU + 16 OMP threads
&
GPU: single-GPU serial; CPU: single-node MPI (OMP=1)
&
Single-node GPU; no MPI/OpenMP; single CUDA kernel
\\

\midrule

\textbf{Workload config} &
\texttt{gmx mdrun}; \texttt{.tpr} inputs; 200k steps; mixed precision; full GPU offload (PME, bonded, nonbonded, update); 0.2\,h runtime limit
&
GPU: \texttt{pmemd.cuda}, \texttt{.mdin}/\texttt{.prmtop}/\texttt{.inpcrd} inputs; explicit solvent; GPU-offloaded PME/forces; \newline
CPU: \texttt{pmemd.MPI}/\texttt{sander.MPI}; MPI-only; walltime controlled
&
CUDA stress kernel; non-MD workload; deterministic FP loop; fixed 60\,s runtime
\\

\midrule

\textbf{Software stack} &
GROMACS 2024.4; GCC 11.2 + MKL + CUDA; SLURM
&
AMBER 24 (GPU), AMBER 20 (CPU); CUDA \texttt{pmemd.cuda}; Intel MPI/OpenMPI
&
FIRESTARTER CUDA build; CUDA 12.9; standalone execution
\\
\hline
\end{tabular}

\end{threeparttable}
\end{table*}
\textit{\textbf{Open-source dataset artifact}}\quad
All data required to reproduce this work are available in our performance–power–energy artifact repository\footnote{\url{https://github.com/AyeshaAfzal91/GPU-Energy-Sweetspot}\label{foot:AD}}; Zenodo DOI to be added upon acceptance. The repository contains datasets, input files, machine-state files documenting hardware and software environments, and scripts for experimental design, methodology, and figure generation. Results (including runtime, power logs, and derived metrics) are stored in uniquely named directories indexed by GPU model, job ID, and clock or power-cap settings to enable fully reproducible energy-efficiency analysis across architectures.

\section{Analytical power-frequency modeling}\label{sec:powermodel}
\begin{figure}[t]
    \centering
        \includegraphics[width=\linewidth]{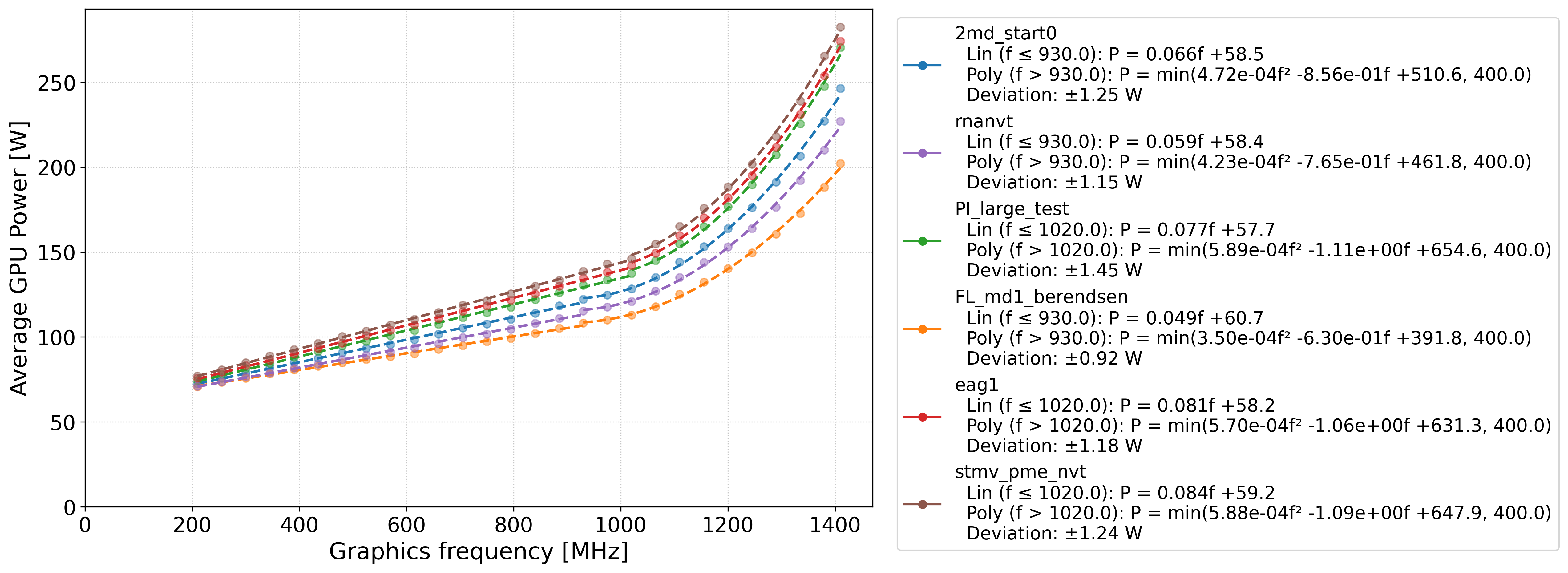}%

    \caption{Piecewise power--frequency fitted model of five GROMACS GPU workloads on H100. Plots for the remaining workloads and GPUs are available at [\href{https://github.com/AyeshaAfzal91/GPU-Energy-Sweetspot/blob/main/PaperFigures/Fig3-All-GPUs-Workloads.pdf}{Link}].
    }    
    \label{fig:freq}
\end{figure}

The dynamic power draw of a compute device is commonly 
approximated by
\begin{equation}
W(f) \propto C \cdot V(f)^2 \cdot f,
\end{equation}
where $f$ is the clock frequency and $V(f)$ is the supply voltage~\cite{Horowitz:2005,Weste:2015}. In general, one expects $V(f)\propto f$, which would lead to cubic power-frequency behavior. Current designs show substantial deviations from this ideal due to leakage currents, hard lower limits for the supply voltage, and ``smart'' power 
management algorithms baked into the hardware~\cite{Li:2009,Leng:2013}.
Empirical measurements from MD and stress-test workloads show that GPU power cannot be described by a single smooth law over the full frequency range; instead, distinct regimes exist; see Fig.~\ref{fig:freq}.
We model GPU power consumption using a piecewise function:
\begin{equation}
W(f) =
\begin{cases}
W_0 + b_1 f, & f < f_t \\
W_0 + a_1 f + a_2(f-f_t)^2, & f \geq f_t
\end{cases}
\label{eq:piecewise}
\end{equation}
Here, $W_0$ denotes the baseline and leakage-dominated power, $b_1$ is the linear dynamic power scaling, $a_1$ the residual linear scaling at high frequency, $a_2$ the nonlinear high-frequency voltage-frequency and thermal scaling, 
and $f_t$ is the transition frequency between linear and nonlinear regimes.
The fitted coefficients are obtained through nonlinear least-squares regression on measured $(f,W)$ samples for each workload and architecture combination (see below). The model is not necessarily continuous at $f=f_t$, but
in general we observe $b_1\approx a_1$.
The transition frequency $f_t$ marks the onset of superlinear power growth. Its location depends on architecture and workload characteristics, particularly compute-bound and memory-bound behavior.
In the linear power scaling regime $f<f_t$, increasing the frequency is energetically favorable as long as $W_0>0$ if we assume that the code performance is proportional to $f$.  This is because the runtime decreases
faster than the power increases. In the quadratic regime $f\geq f_t$,
power grows faster than performance, leading to diminishing energy efficiency. 
Assuming near-linear performance scaling, $P(f)\propto f$, energy efficiency in this regime becomes
\[
\eta(f)=\frac{P(f)}{W(f)}
\propto
\frac{f}{W_0+a_1f+a_2(f-f_t)^2}
\quad (f>f_t).
\]
The stationary condition $\frac{d\eta}{df}=0$ yields the well-known approximate optimum
\begin{equation}
f^\ast \approx \sqrt{\frac{W_0}{a_2}},
\label{eq:fopt}
\end{equation}
which is independent of $a_1$~\cite{AfzalThesis:2015,Hager:2016}.
This shows that a large base power $W_0$ shifts the optimum toward higher frequencies, whereas a larger $a_2$ shifts it toward lower frequencies.
This analysis also shows that, under the assumptions used to derive Eq.~\eqref{eq:fopt}, the optimal frequency $f^\ast$ cannot be smaller than $f_t$.

\begin{figure}[t]
    \centering
    \subfloat[Frequency operating landscape]{%
        \includegraphics[width=0.48\linewidth]{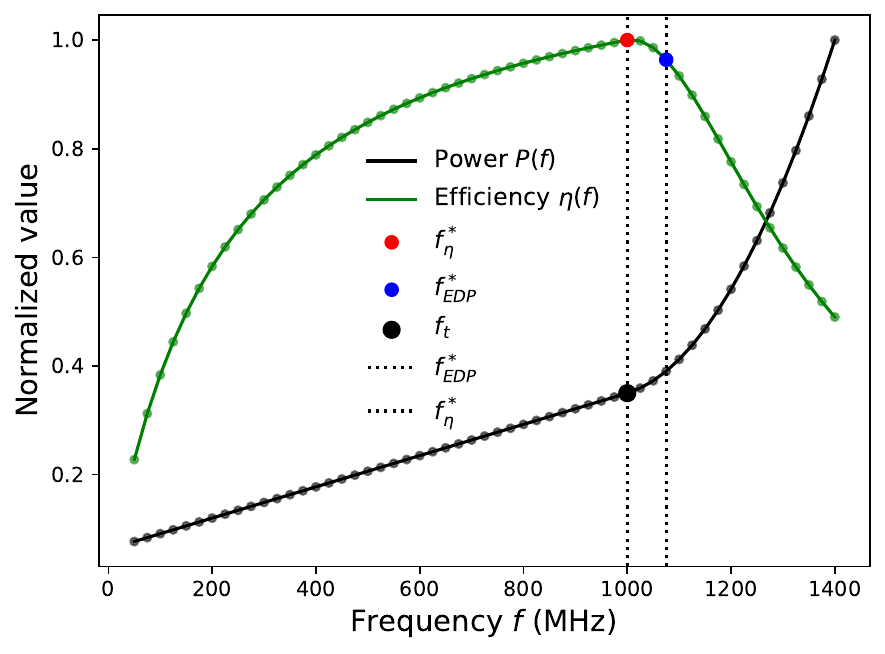}%
    }\hfill
    \subfloat[Optimal frequency phase space]{%
        \includegraphics[width=0.48\linewidth]{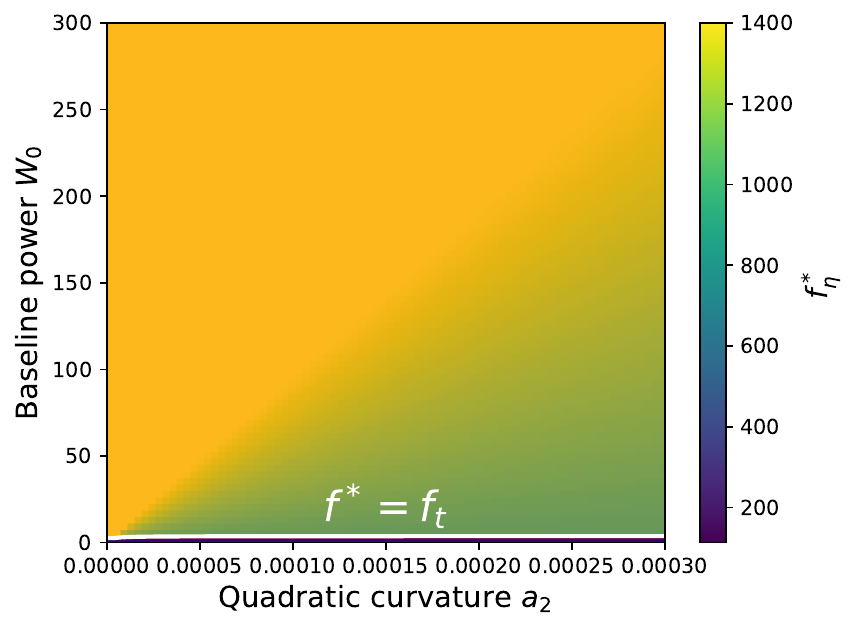}%
    }
    \caption{Piecewise GPU power–frequency model and energy-optimal regime structure. (a) Normalized power $P(f)$ (black) and normalized energy efficiency $\eta(f)$ (green) vs. frequency, showing optima ($f^\ast_\eta$ in red, $f^\ast_{\mathrm{EDP}}$ in blue) and transition point $f_t$ in black. (b) Phase diagram of $f^\ast_\eta$ as a function of baseline power $P_0$ and quadratic curvature $a_2$. The color map shows numerically obtained efficiency-optimal frequency $f^\ast_\eta$ (Eq.~\ref{eq:fopt}), from high (yellow) to low (blue). Shaded regions indicate whether the optimum lies in the linear or quadratic power regime, with the horizontal boundary at $f^\ast_\eta = f_t$.}
    \label{fig:modelimp}
\end{figure}
\textit{\textbf{Model implications}}\quad
The model provides a useful interpretation of GPU frequency scaling behavior: (1) efficiency is mainly governed by regime transitions, not smooth scaling, and (2) $W_0$ and $a_2$ dominate optimal operation. In Fig.~\ref{fig:modelimp}(a) we show a typical scenario for a compute-bound
code (i.e., performance is linear in the clock speed) with  
a piecewise linear-to-quadratic GPU power-frequency relation. 
The model reproduces the DVFS landscape observed empirically (see Fig.~\ref{fig:freq} and Fig.~\ref{fig:Zplot} for real data).
It can be used to identify two distinct optima: the energy-efficiency maximum at $f^*_{\eta}$ and the EDP minimum at $f^*_{\mathrm{EDP}}$. For compute-bound code, the former is always at or to the right of the frequency
transition point $f_t$.
The phase diagram (Fig.~\ref{fig:modelimp}(b)) explores the dependence of $f^\ast$ on baseline power $W_0$ and quadratic curvature $a_2$, confirming Eq.~\ref{eq:fopt}. Increasing $W_0$ shifts $f^\ast$ upward (toward yellow), as higher performance is needed to amortize fixed baseline power, whereas increasing $a_2$ shifts $f^\ast$ downward (toward blue) due to stronger superlinear growth in the high-frequency regime. The linear term $a_1$ is negligible for compute-bound workloads but can become relevant when memory-bound effects distort performance linearity, causing increased idle cycles in the cores. A boundary at $f^\ast=f_t$ separates linear- (bottom) and curvature-dominated (top) regimes, illustrating that energy-optimal GPU operation is a multi-parameter, regime-based problem.

\begin{figure}[t]
    \centering
    \subfloat[GPU MD workloads]{%
        \includegraphics[width=0.374\textwidth]{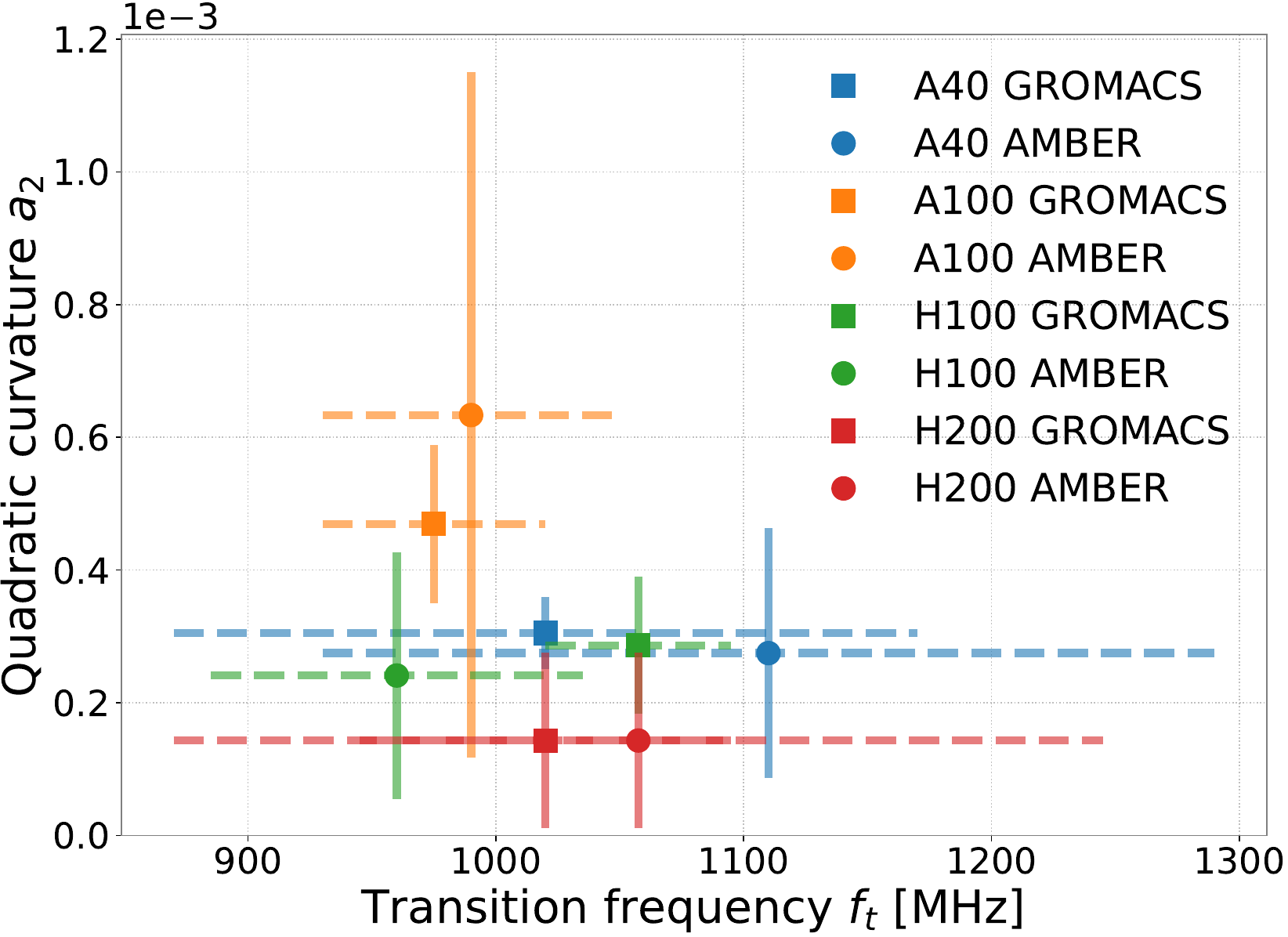} %
    } \quad
    \subfloat[CPU MD workloads]{%
    \includegraphics[width=0.39\linewidth]{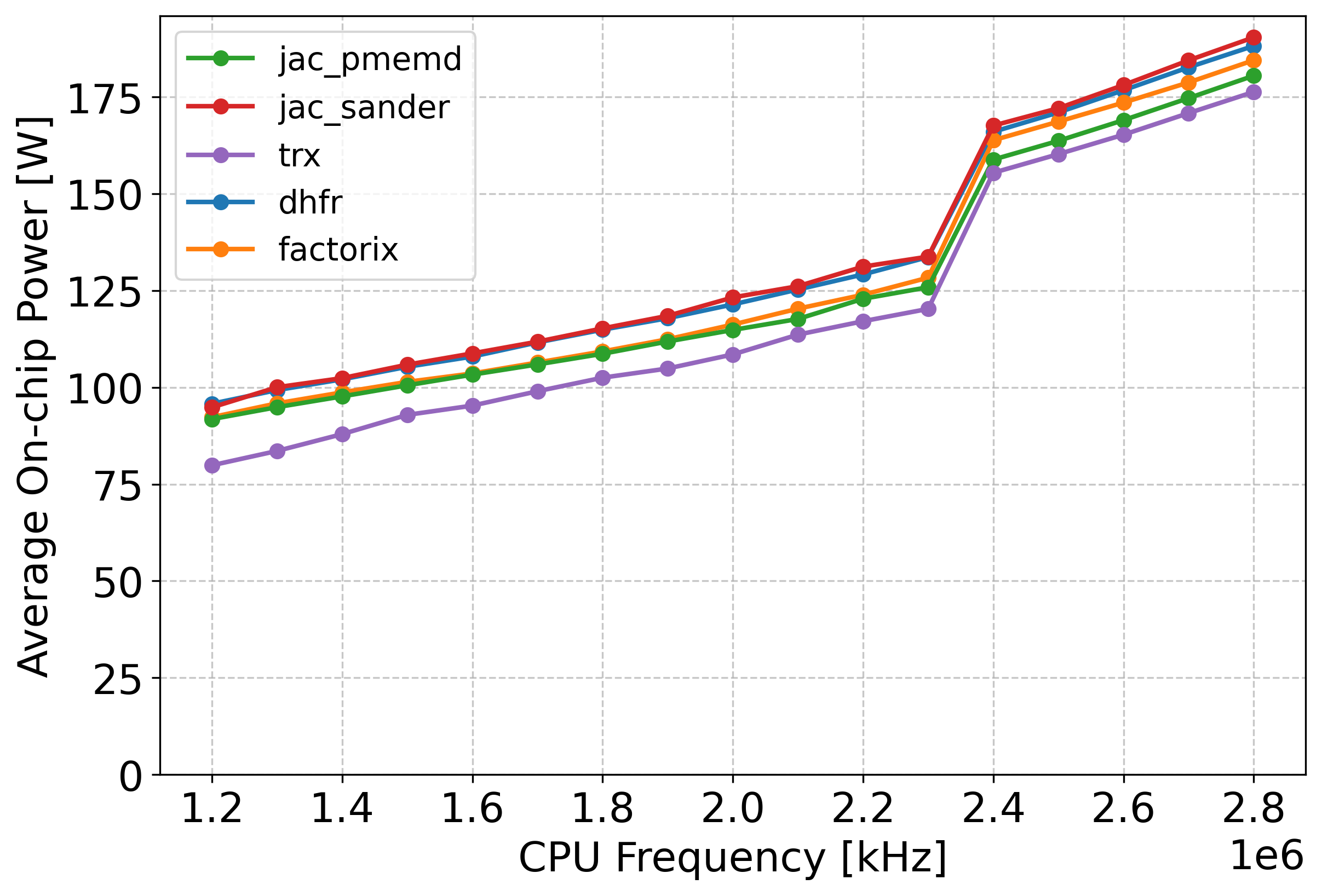}
    } 
    \caption{(a) Ranges of high-frequency GPU power curvature $a_2$ (vertical dash lines) and transition frequency $f_t$ (horizontal dotted lines), with workload midpoints as markers (circle: AMBER, square: GROMACS) and colors indicating GPU models.
    (b) CPU power–frequency scaling on Intel Ice Lake for AMBER, mostly linear ($a_2=0$).}
    \label{fig:ft_a2}
\end{figure}

\textit{\textbf{Fitting to GPU workloads}}\quad
GPU power is modeled using bench\-mark-specific piecewise regression on $(f,W)$ data. The legend in Fig.~\ref{fig:freq} shows model coefficients and errors for all six GROMACS benchmarks on H100. The ranges of transition frequency $f_t$ and corresponding quadratic curvature $a_2$ across GPUs and workloads are summarized in Fig.~\ref{fig:ft_a2}(a).  
The transition point $f_t$ is identified via exhaustive search over interior indices of each frequency sweep, splitting the low- and high-frequency regimes. Observed $f_t$ ranges with AMBER and GROMACS workloads exhibit distinct sub-ranges within each GPU architecture.
For each candidate split, a linear model is fitted to the low-frequency regime and a quadratic model to the high-frequency regime. Parameters are estimated via nonlinear least squares (Levenberg--Marquardt), and the optimal split minimizes the combined sum of squared errors:
$
\mathrm{SSE}_{\mathrm{total}} = \mathrm{SSE}_{\mathrm{lin}} + \mathrm{SSE}_{\mathrm{quad}}.
$
Fits are discarded if convergence fails or if fewer than six samples are available per segment. To prevent unphysical extrapolation, the quadratic model is optionally compared against the maximum power envelope. Model accuracy is quantified using the root-mean-square error (RMSE), computed from the total residual error of the selected piecewise fit:
\begin{equation}
  \mathrm{RMSE} = \sqrt{\frac{\mathrm{SSE}_{\mathrm{lin}} + \mathrm{SSE}_{\mathrm{quad}}}{N}},
\end{equation}
where $N$ is the total number of samples and the errors are computed from the total residual error at the optimal split. Across benchmarks and architectures, RMSE remains within a few Watts (typically $<1$\% relative error, with maximum deviations of $<1.8$\% only for large benchmarks on H100 due to smooth saturation effects), confirming that the piecewise formulation with its linear and quadratic branches is a useful approximation.
Fig.~\ref{fig:ft_a2}(a) shows clear trends: old-generation GPUs exhibit higher $a_2$, indicating stronger superlinear power growth at high frequencies than later-generation GPUs. Transition frequencies vary widely across workloads and $a_2$ is always positive, implying that high-frequency nonlinear effects may slightly amplify or suppress power depending on workload and GPU.

\textit{\textbf{Comparison with CPU power-frequency behavior}}\quad
Modern CPU power consumption exhibits near-linear dependence on frequency as shown by the five compute-bound AMBER benchmarks on an Intel Ice Lake CPU across the evaluated DVFS range, as shown in Fig.~\ref{fig:ft_a2}(b). No pronounced superlinear regime is observed, indicating negligible DVFS curvature and weak thermal or voltage-frequency nonlinearities under workloads.
CPU power is modeled as:
\begin{equation}
P_{\mathrm{CPU}}(f) = W_0^{c} + \alpha f + \Delta P_u(f_u),
\end{equation}
where $W_0^{c}$ is the static baseline power, $\alpha$ the linear dynamic scaling coefficient. $\Delta P_u$ is a discrete offset induced by the uncore frequency $f_u$ which vanishes when the uncore frequency is fixed, reducing the model to a purely linear form.
A reproducible power jump occurs at $f_u = 2.3$\,GHz (Fritz). This power jump is consistent across all workloads and disappears when the uncore frequency is fixed~\cite{AfzalHW:2025:1}.
Since CPU DVFS power is dominated by uniform switching activity rather than workload-dependent nonlinear effects, CPU energy-to-solution is weakly sensitive to frequency scaling.

\section{Energy-efficiency sweet spots}\label{sec:eval} 
In this section, we examine how frequency scaling and power caps interact differently with GPU workloads and power regimes, although both aim to reduce energy compared to default operation. Energy efficiency depends primarily on workload intensity relative to GPU power regimes, which can be grouped into three classes: (1) small workloads underutilize GPUs, yielding poor efficiency; (2) medium workloads balance utilization and efficiency; (3) large workloads enter power-limited regimes with diminishing returns. These align with 
Sec.~\ref{sec:powermodel}, where energy degradation begins above the transition frequency $f_t$.

\begin{figure}[tbh!]
    \centering
    \subfloat[A40-AMBER]{%
        \includegraphics[width=0.45\linewidth]{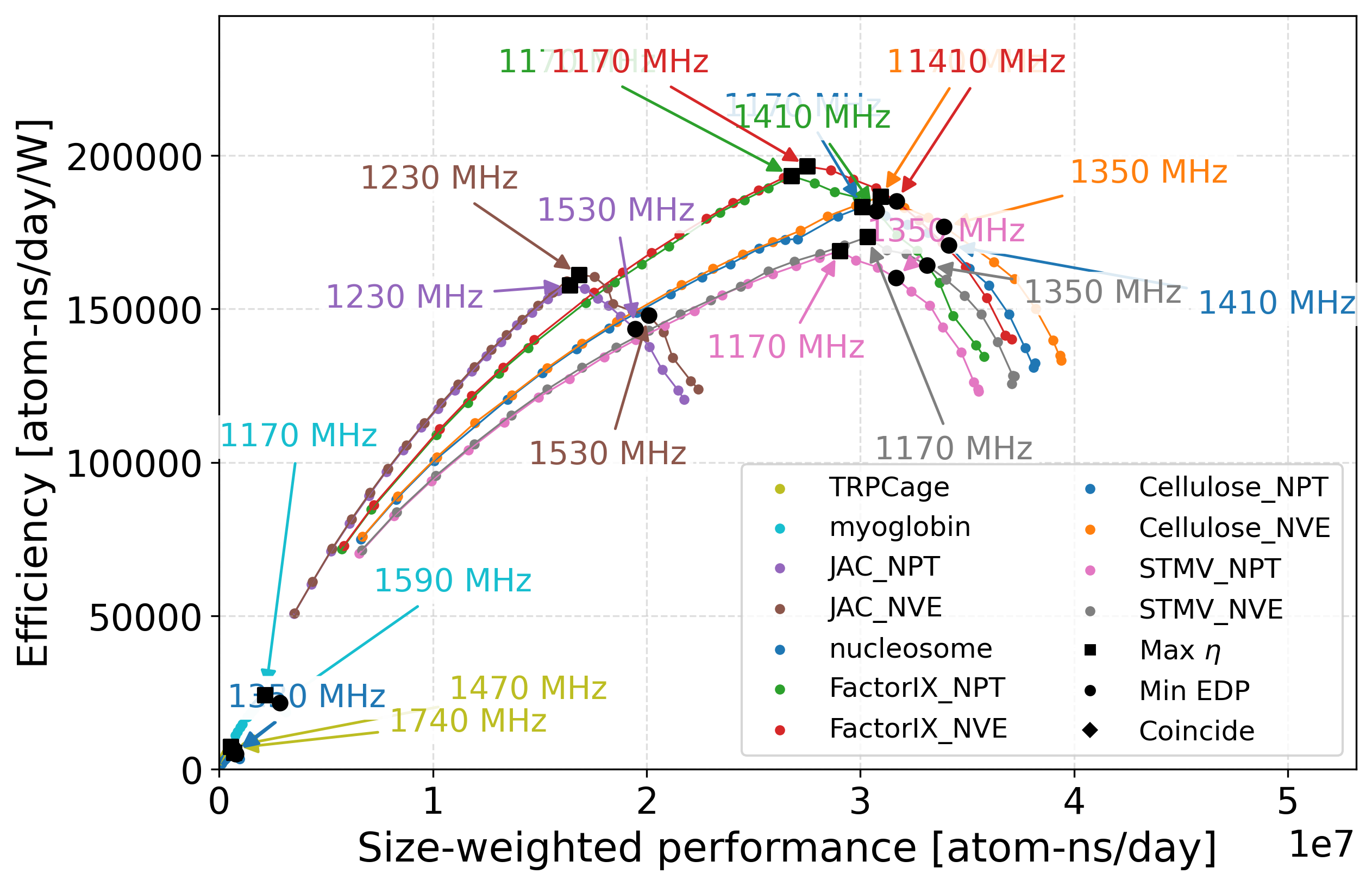}%
    }\quad
    \subfloat[A100-AMBER]{%
        \includegraphics[width=0.45\linewidth]{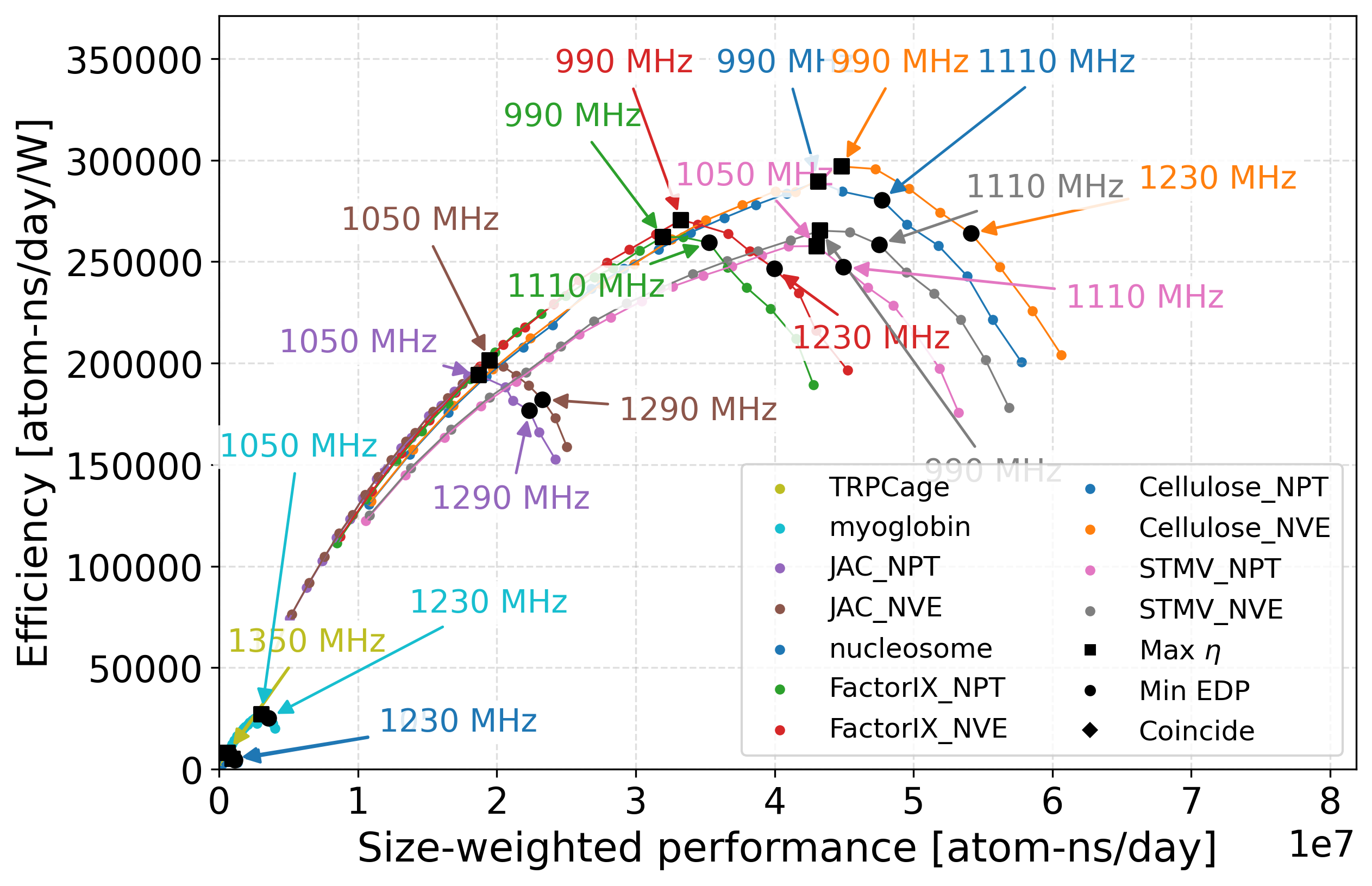}%
    }\hfill
    \subfloat[H100-AMBER]{%
        \includegraphics[width=0.45\linewidth]{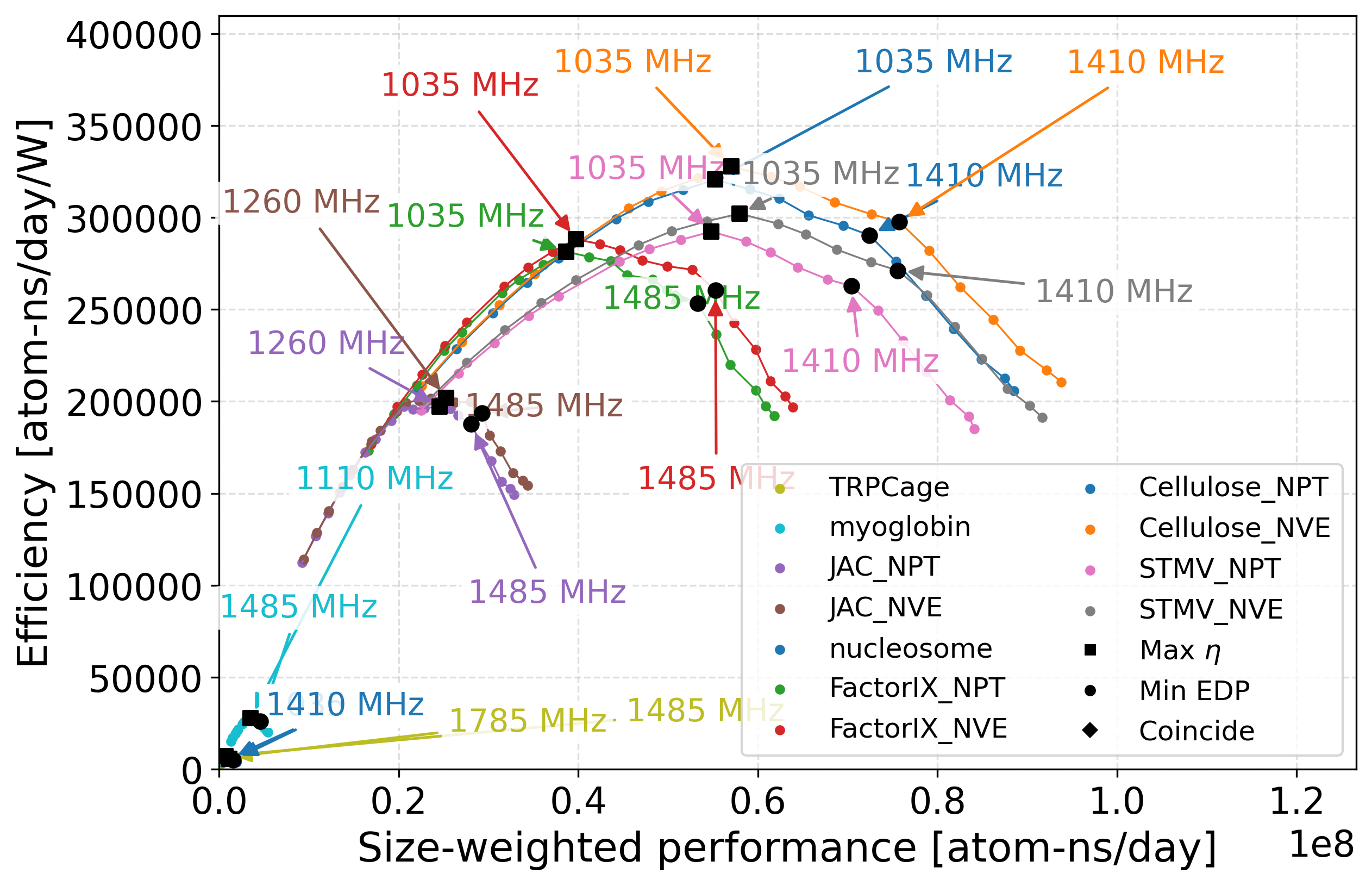}%
    }\quad
    \subfloat[H200-AMBER]{%
        \includegraphics[width=0.45\linewidth]{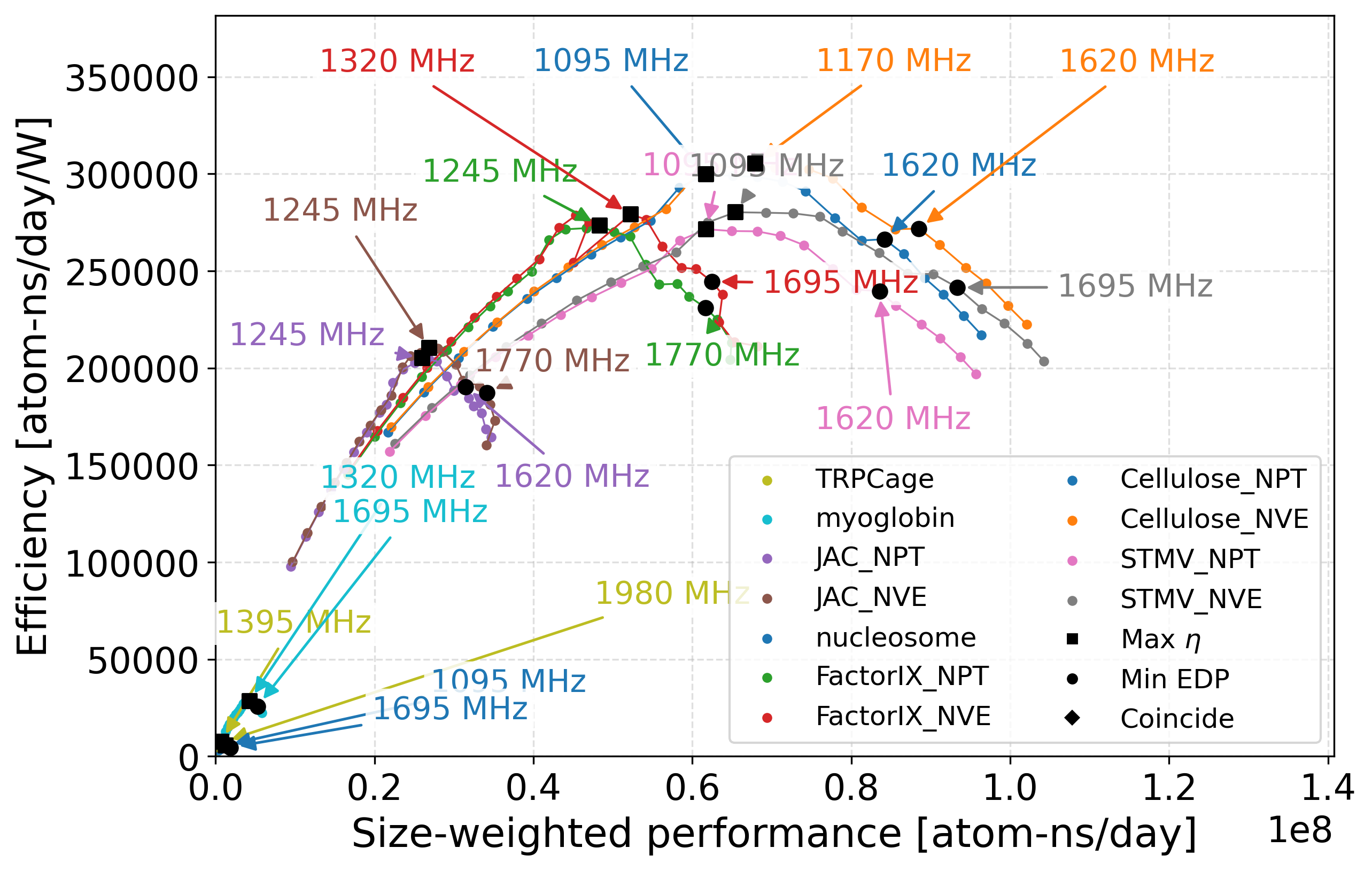}%
    }\hfill
    \subfloat[A40-GROMACS]{%
        \includegraphics[width=0.45\linewidth]{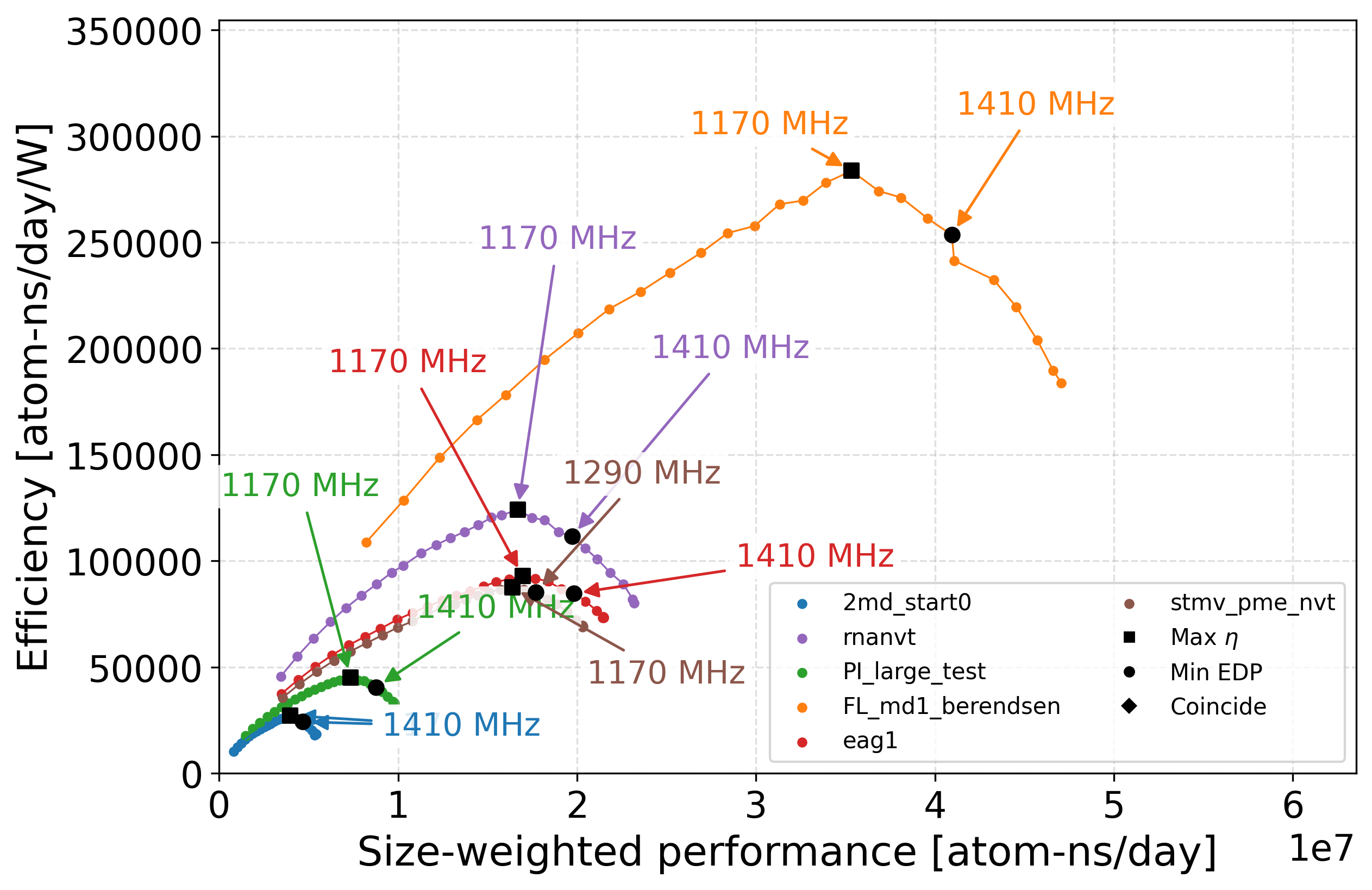}%
    }\quad
    \subfloat[A100-GROMACS]{%
        \includegraphics[width=0.45\linewidth]{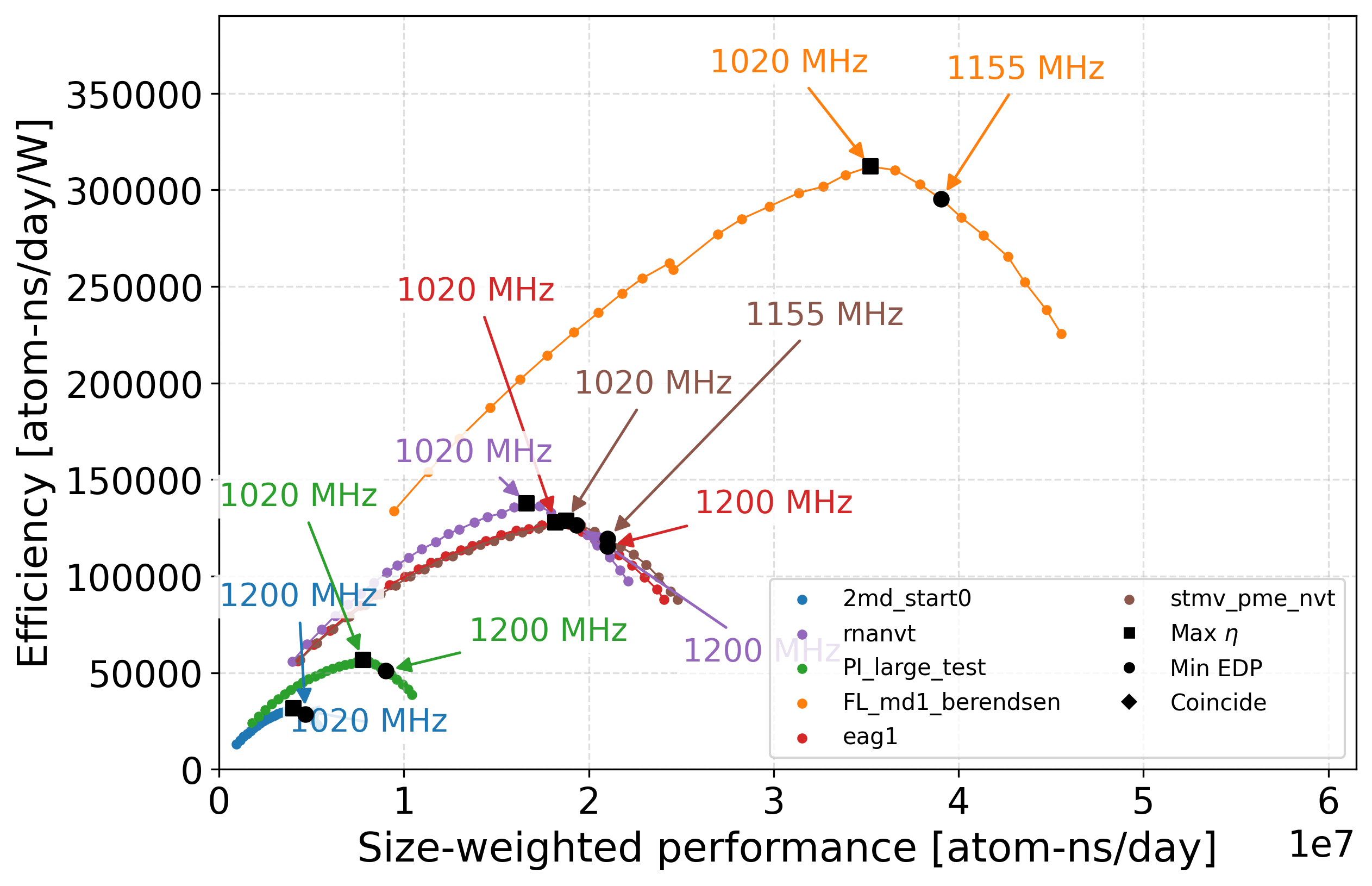}%
    }\hfill
    \subfloat[H100-GROMACS]{%
        \includegraphics[width=0.45\linewidth]{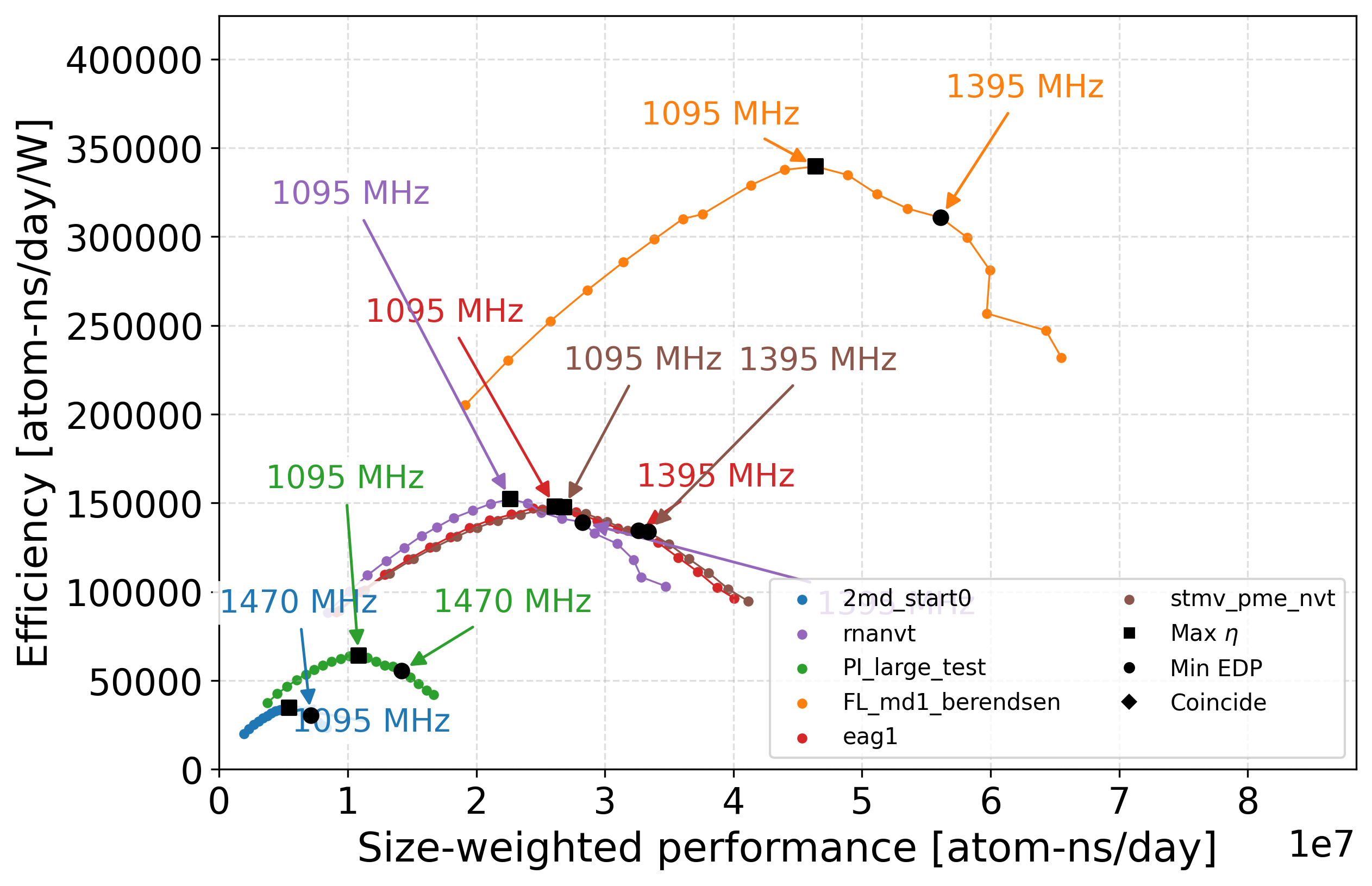}%
    }\quad
    \subfloat[H200-GROMACS]{%
        \includegraphics[width=0.45\linewidth]{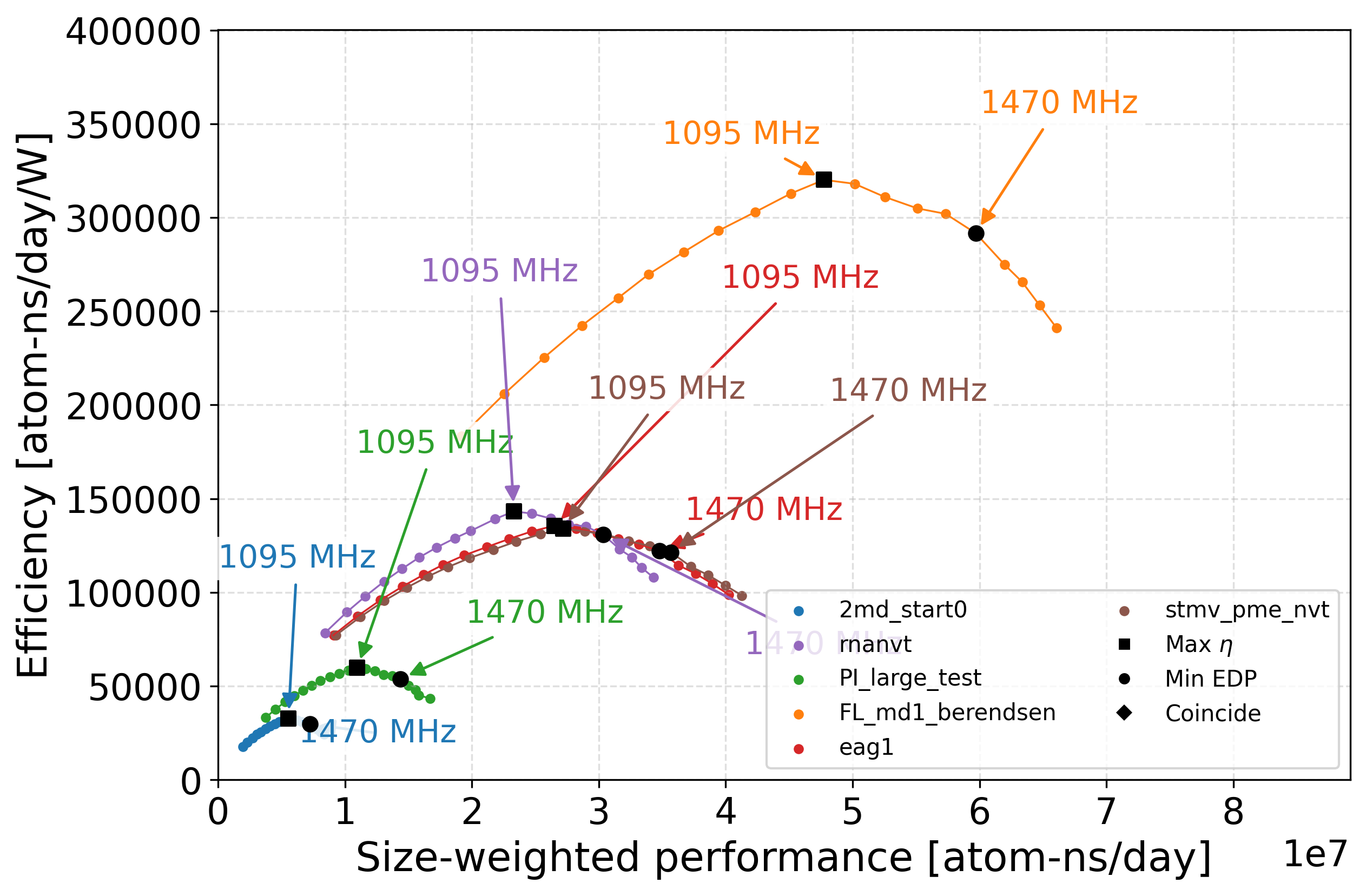}%
    }
    \caption{Performance–energy efficiency Pareto space under frequency scaling. The maximum energy efficiency ($f^\ast_{\eta}$) and minimum EDP ($f^\ast_{\mathrm{EDP}}$) frequencies are labeled. 
    }
    \label{fig:Zplot}
\end{figure}
\begin{figure}[t]
    \centering
        \centering
    \subfloat[AMBER]{%
        \includegraphics[width=0.45\linewidth]{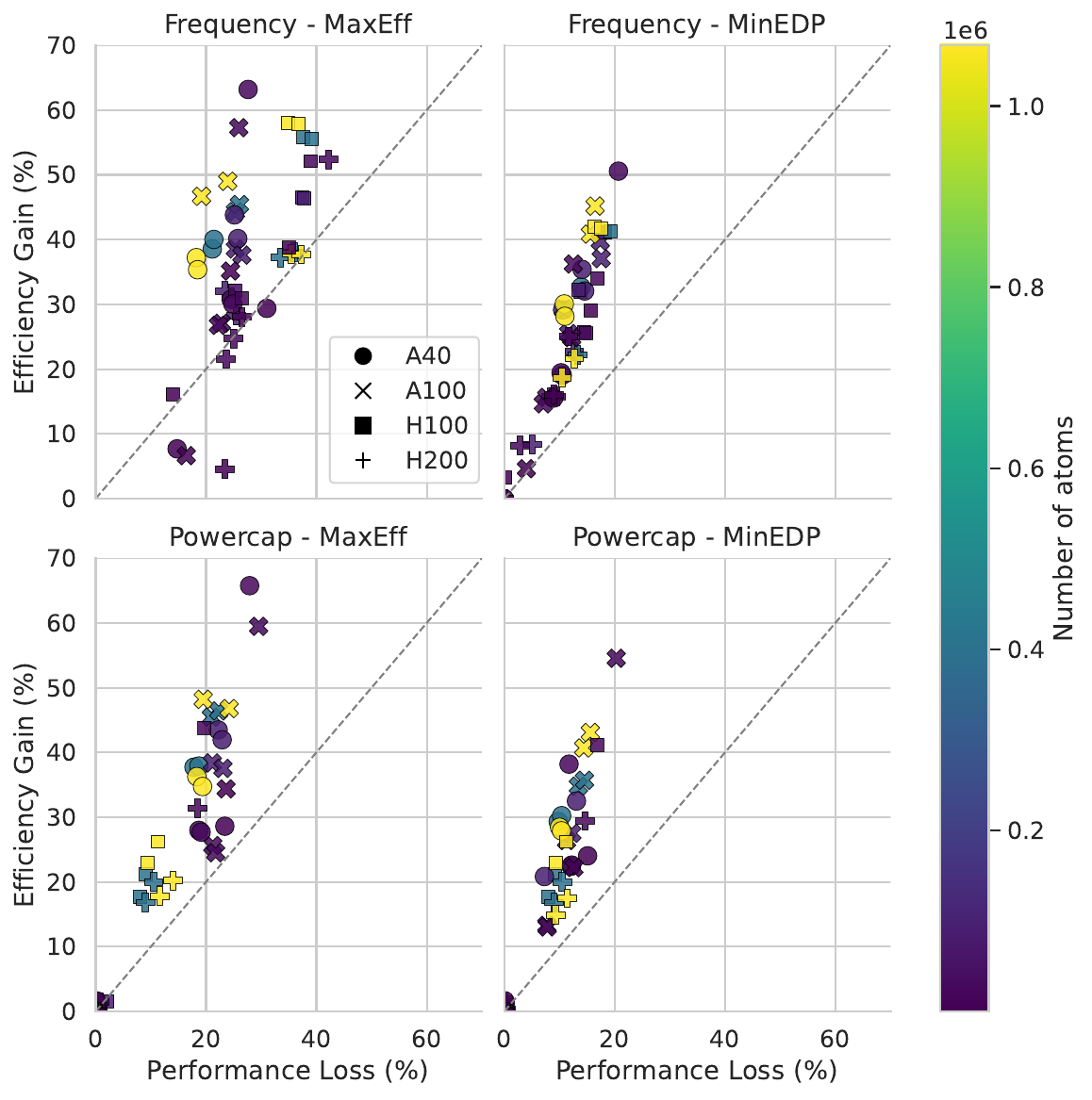}%
    }\quad
    \subfloat[GROMACS]{%
        \includegraphics[width=0.45\linewidth]{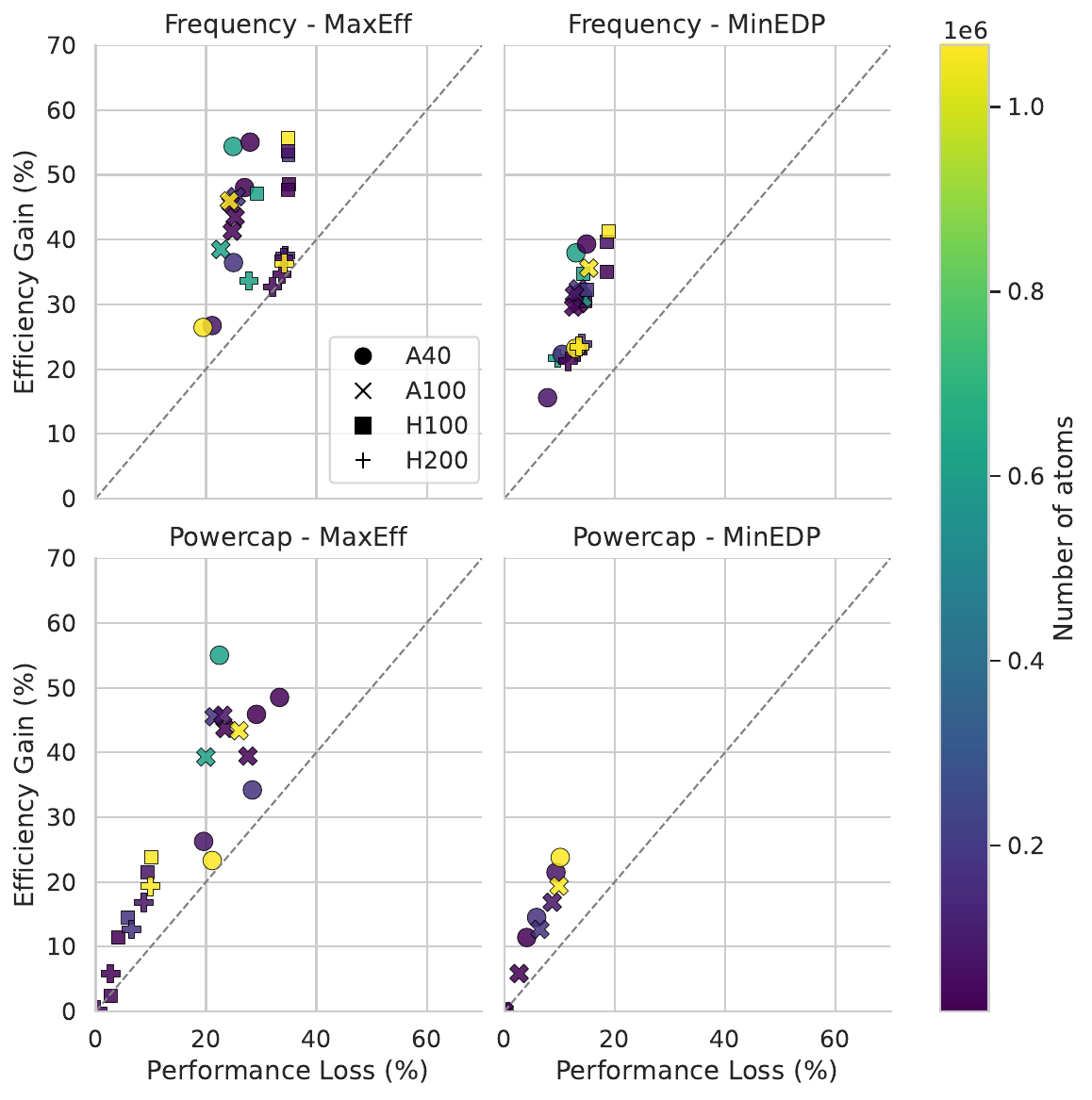}%
    }
\caption{Efficiency--performance trade-off for frequency scaling and power-cap settings at maximum-efficiency ($f^\ast_{\eta}$, $W^\ast_{\eta}$) and minimum-EDP ($f^\ast_{\mathrm{EDP}}$, $W^\ast_{\mathrm{EDP}}$) points. Each point shows efficiency gain ($\Delta \eta$) vs. performance change ($\Delta P$) relative to default. Markers denote GPU type; color encodes workload size from small (blue) to large (yellow).} 

    \label{fig:tradeoff}
\end{figure}

\textit{\textbf{Frequency scaling and energy-efficiency regimes}}\quad
Figure~\ref{fig:Zplot} relates energy efficiency to code performance, with GPU frequency as a parameter. Horizontal lines indicate constant efficiency, vertical lines indicate constant performance, with the peak marking the efficiency optimum $f^*_{\eta}$. Panels (a)--(h) show that frequency scaling exposes a broad energy--performance space across all architectures. Reducing frequency lowers power faster than performance degrades, improving efficiency until the workload-specific optimum $f^*_{\eta}$ is reached. Beyond this point, runtime growth dominates, increasing energy-to-solution.
In (a)--(d), the smaller AMBER benchmark (Red, Factor IX, 90K atoms) generally performs worse than the larger one (Orange, Cellulose, 400K atoms), except on A40, which has the highest $f_t$ (Fig.~\ref{fig:ft_a2}(a)), reflecting architectural differences.
We investigated whether the maximum-efficiency point coincides with the transition frequency $f_t$. Results show that $f^*_{\eta}$ (black) typically clusters near $f_t$ but does not necessarily coincide. The transition frequency marks the onset of nonlinear power growth, while the quadratic curvature $a_2$ in Eq.~\ref{eq:piecewise} governs efficiency collapse. Thus, $f_t$ acts as a first-order indicator of regime transition, while $a_2$ controls the rate of efficiency deterioration, consistent with Eq.~\ref{eq:fopt}, where higher curvature $a_2$ on early generation A40/A100 GPUs shifts the optimum to lower frequencies.
For compute-bound workloads, the optima remain close to the transition frequency $f_t$ due to near-linear performance scaling (e.g., $f^*_{\eta}\approx f_{t} = 1005$\,MHz for FIRESTARTER; Fig.~\ref{fig:tradeoff}). Few MD workloads show $f^*_{\eta}$--$f_t$ separation, indicating partial memory-bound execution. 
Frequency scaling thus enables systematic navigation of GPU regimes, moving workloads between efficiency-improving, balanced, and energy-inefficient regions of the Pareto frontier.

\begin{figure}[t]
    \centering
    \subfloat[A40-AMBER]{%
        \includegraphics[width=0.32\linewidth]{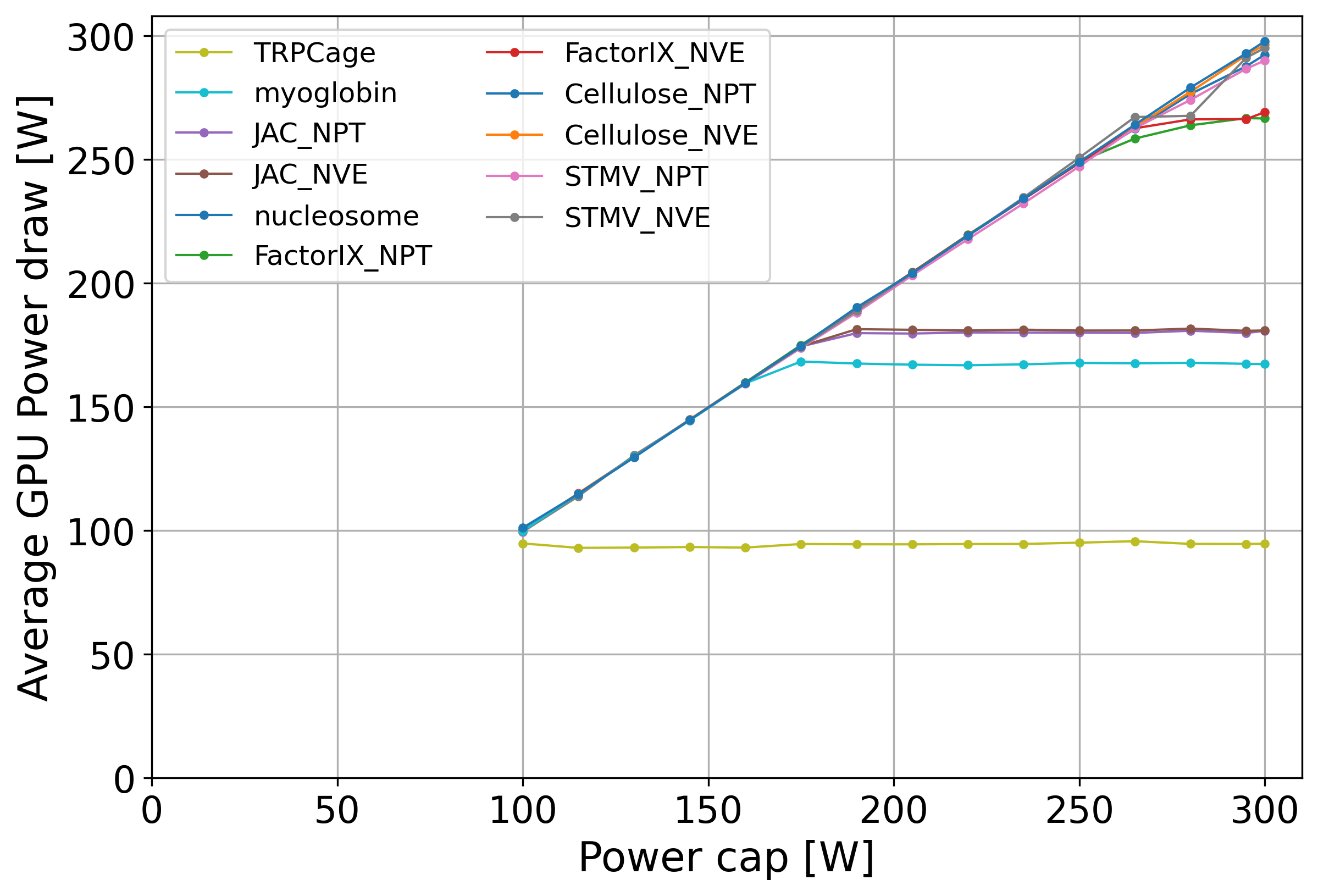}%
    }\hspace{0.1cm}
    \subfloat[A100-AMBER]{%
        \includegraphics[width=0.32\linewidth]{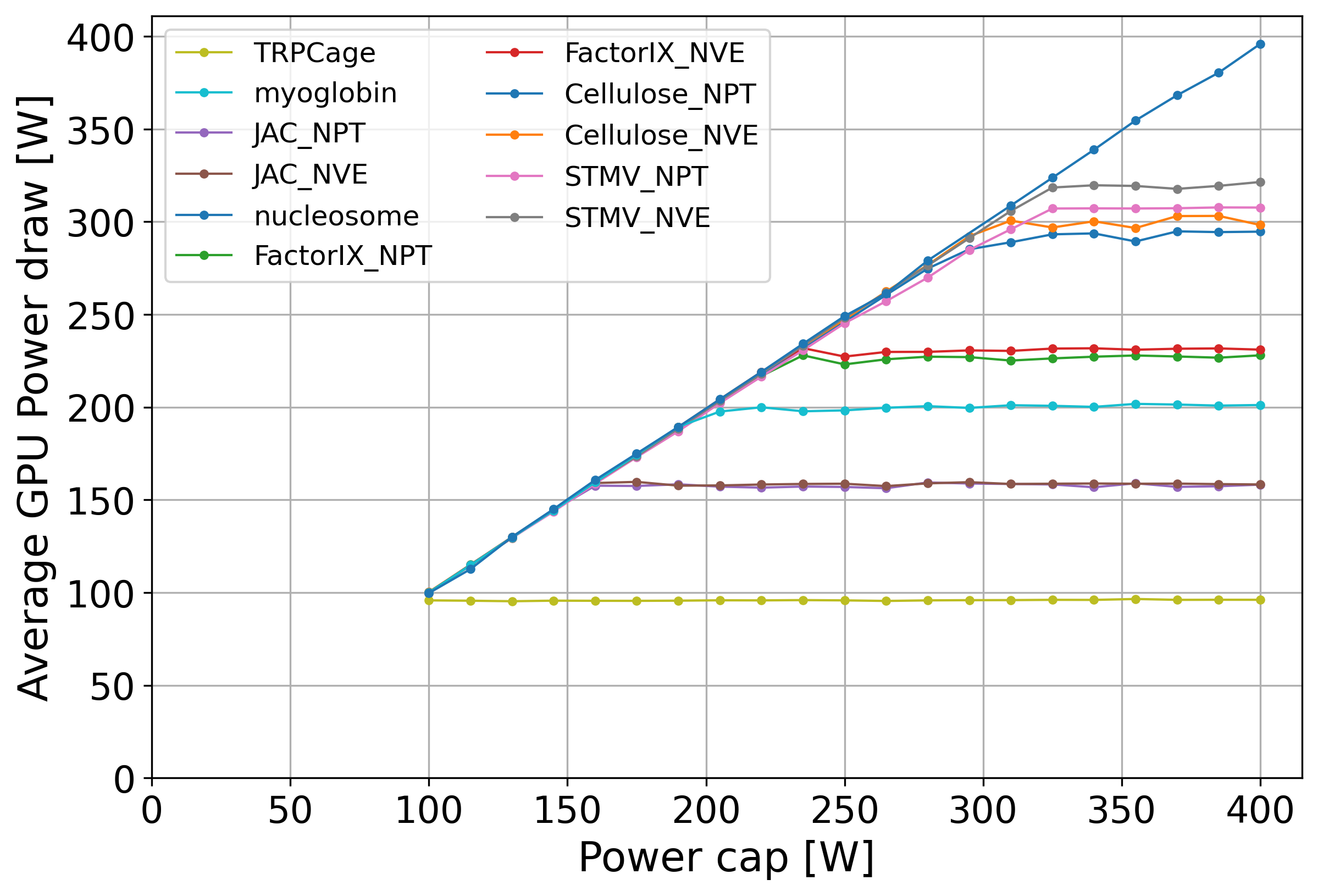}%
    }\hspace{0.1cm}
    \subfloat[H100-AMBER]{%
        \includegraphics[width=0.32\linewidth]{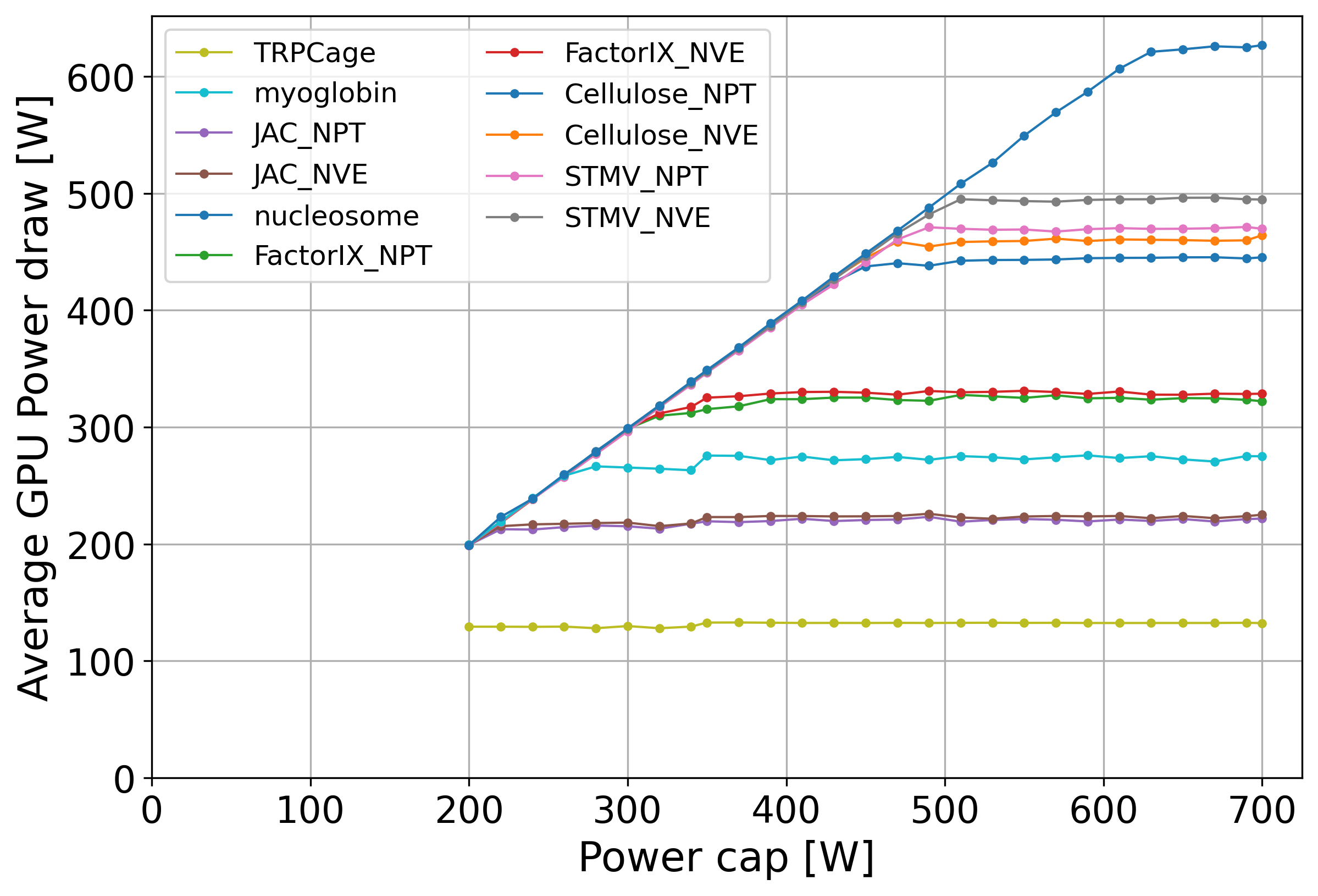}%
    }\hfill
    \subfloat[H200-AMBER]{%
        \includegraphics[width=0.32\linewidth]{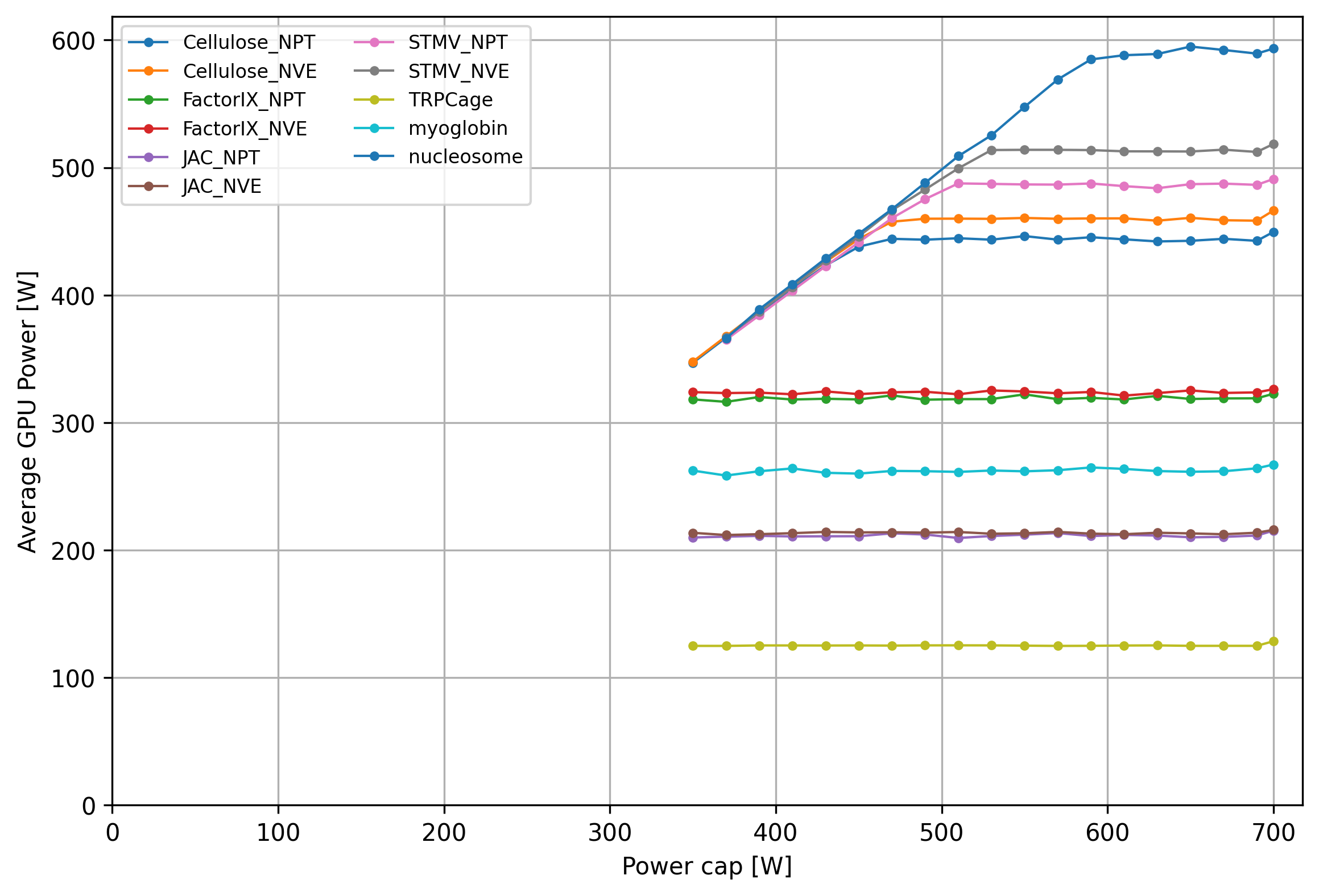}%
    }\hspace{0.1cm}
    \subfloat[A40-GROMACS]{%
        \includegraphics[width=0.32\linewidth]{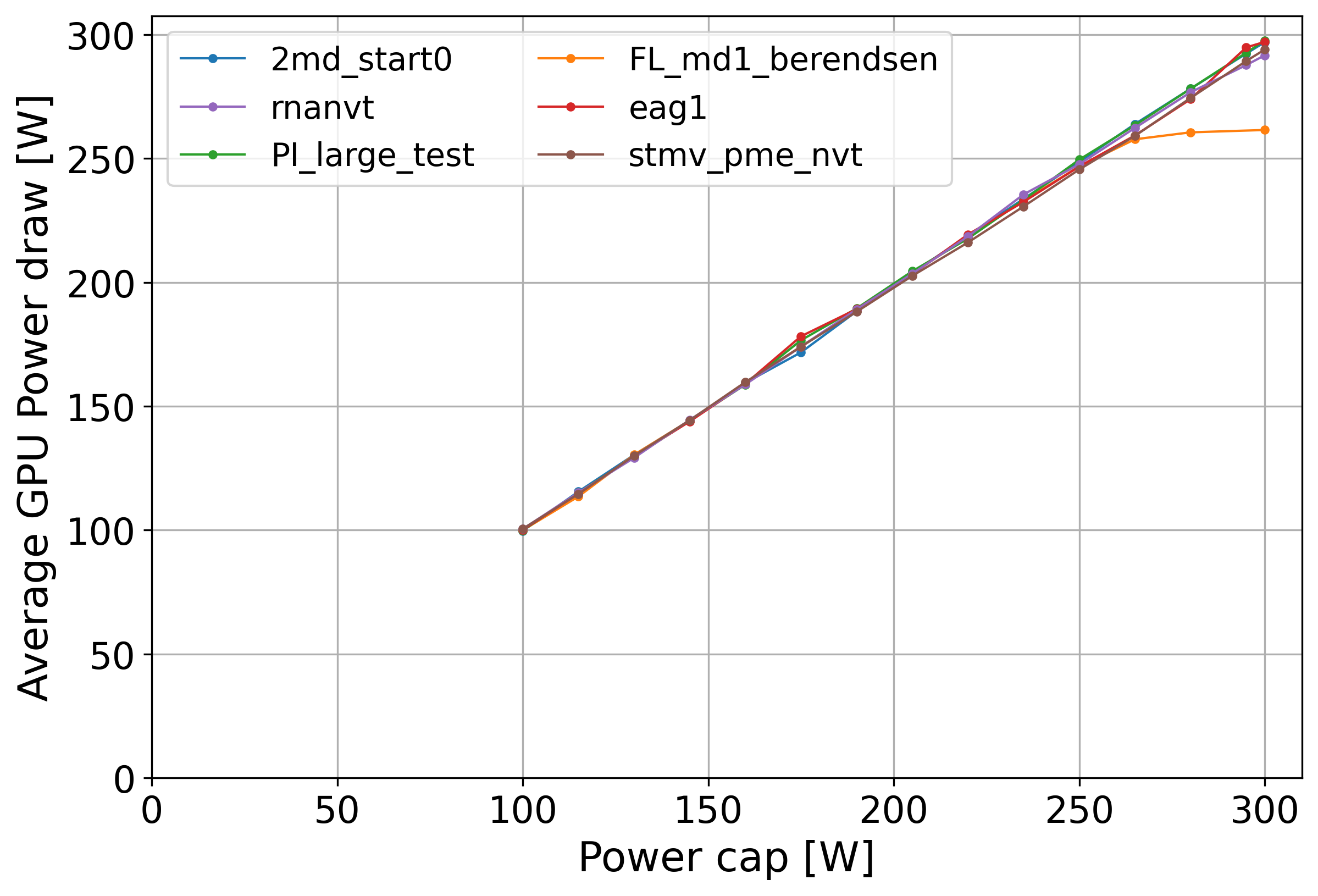}%
    }\hspace{0.1cm}
    \subfloat[A100-GROMACS]{%
        \includegraphics[width=0.32\linewidth]{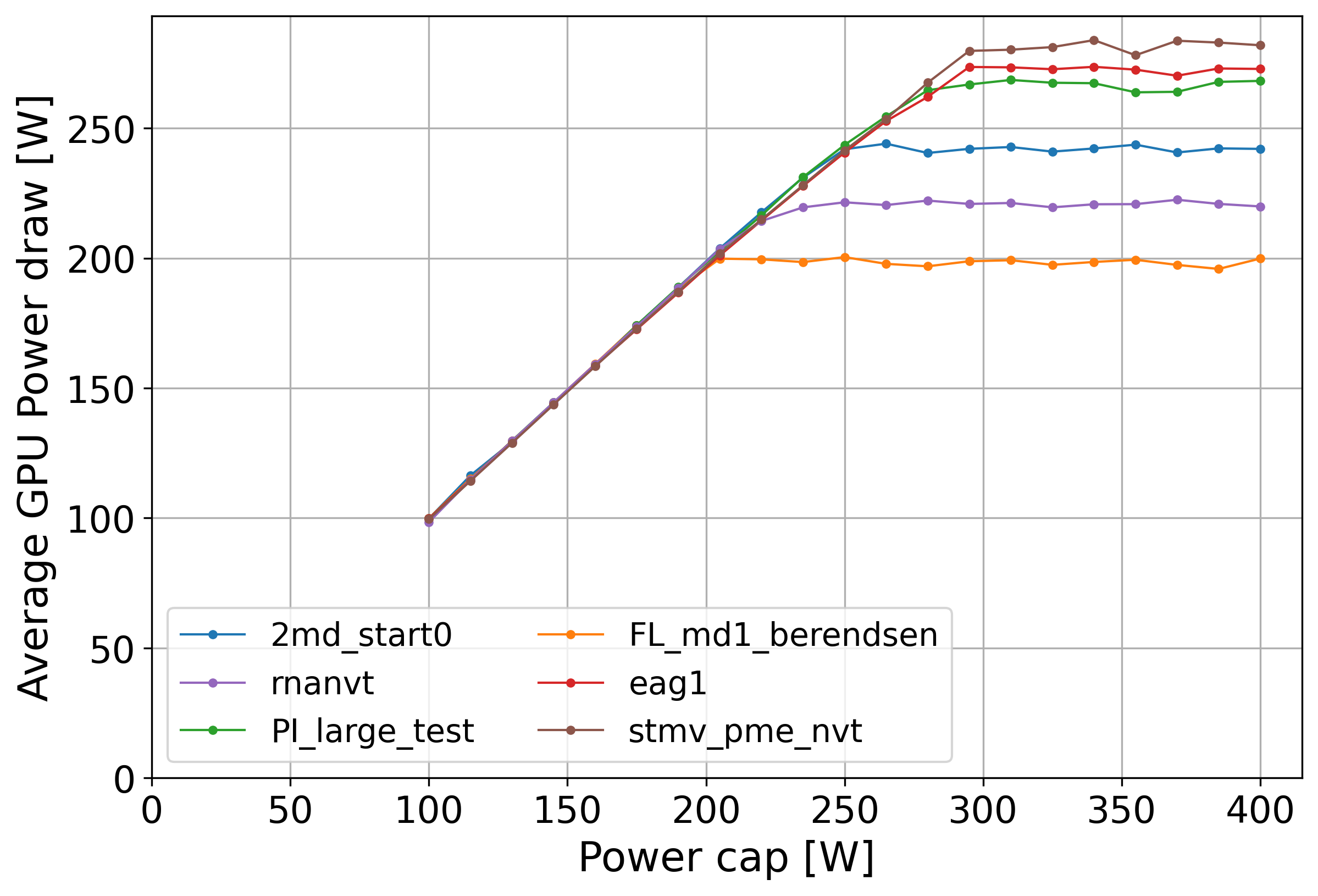}%
    }\hfill
    \subfloat[H100-GROMACS]{%
        \includegraphics[width=0.32\linewidth]{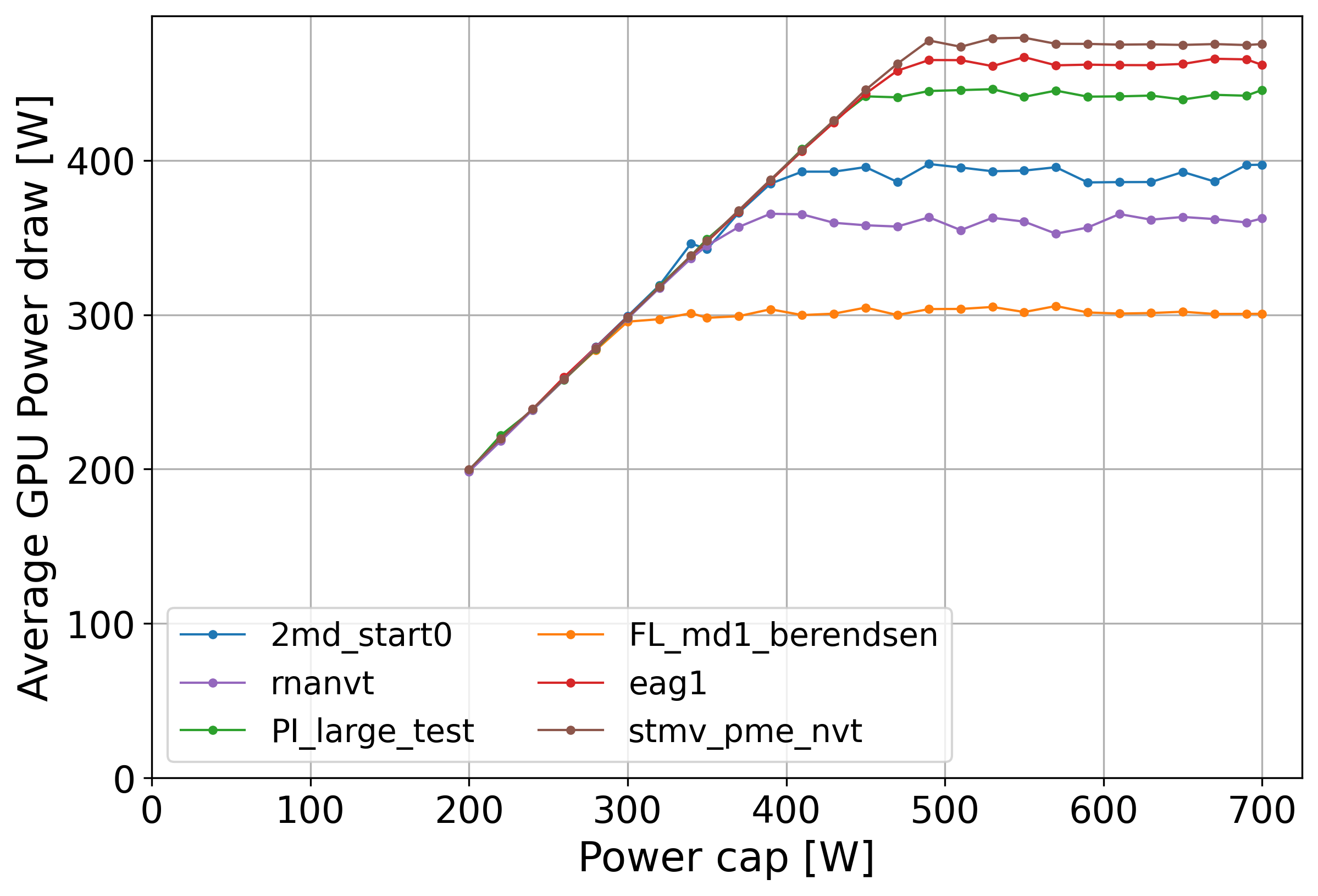}%
    }\hspace{0.1cm}
    \subfloat[H200-GROMACS]{%
        \includegraphics[width=0.32\linewidth]{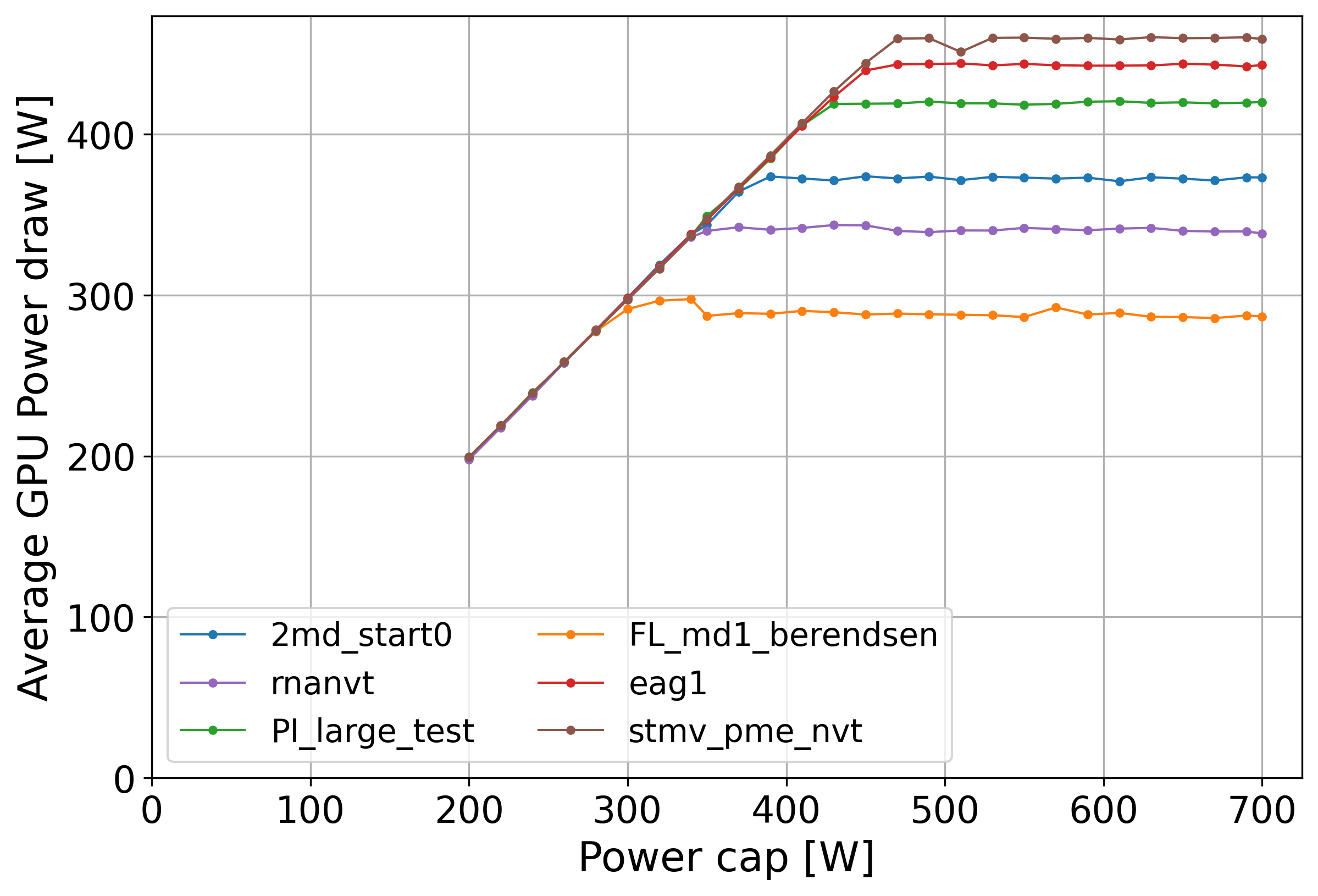}%
    }\hspace{0.1cm}
    \subfloat[FIRESTARTER]{%
        \includegraphics[width=0.32\linewidth]{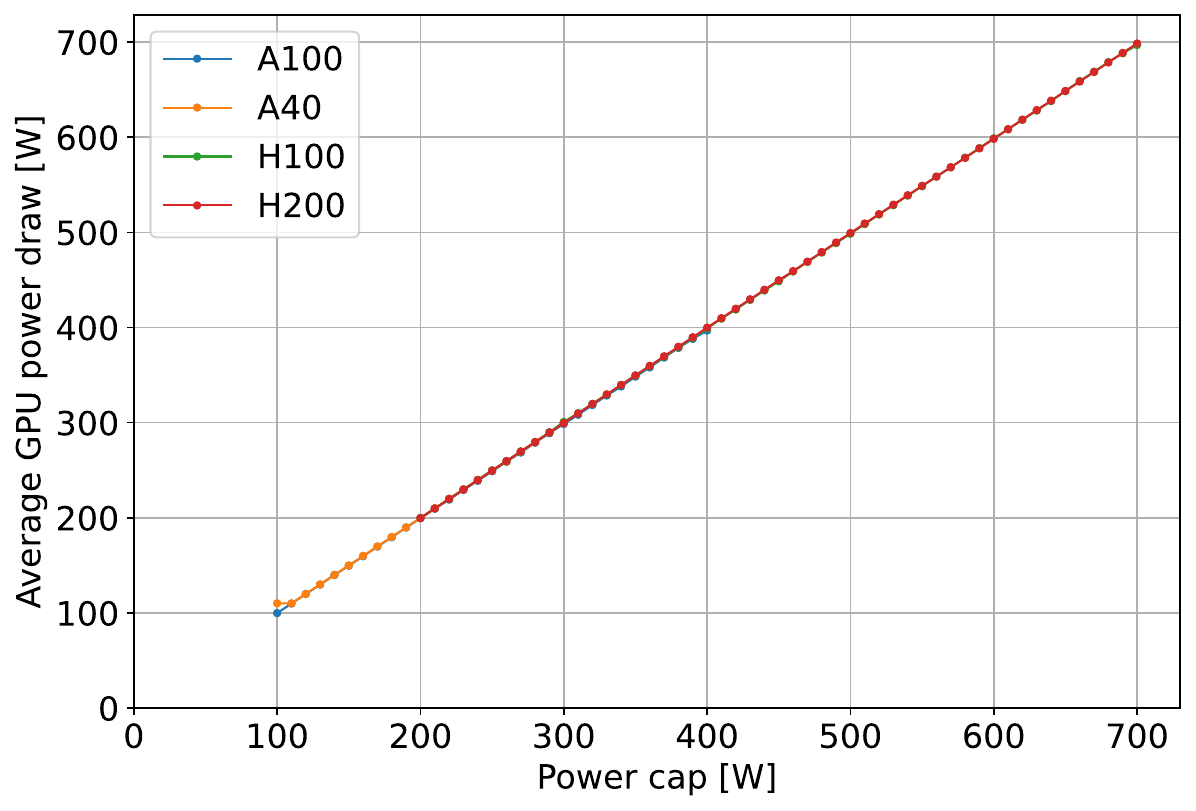}%
    }
    \caption{Impact of hardware power capping on average GPU power across architectures and workloads: (a)–(d) AMBER, (e)–(h) GROMACS, and (i) FIRESTARTER synthetic stress-test baseline on A40, A100, H100, and H200 GPUs.}
    \label{fig:power-powercap}
\end{figure}
\textit{\textbf{Power cap and energy-efficiency regimes}}\quad
The effectiveness of a power cap depends on the interaction between the hardware limit and the workload's natural power demand. Figure~\ref{fig:power-powercap} shows average GPU power as a function of the enforced cap for A40, A100, H100, and H200 GPUs across AMBER, GROMACS, and FIRESTARTER workloads.
A linear increase in measured power with increasing cap, most clearly seen for FIRESTARTER (Fig.~\ref{fig:power-powercap}(i)), indicates continuous hardware throttling, i.e., the workload remains power-limited across the tested range. In contrast, MD applications in panels (a)–(h) exhibit workload-dependent saturation, where power saturates, showing that the cap does not constrain the workload. GROMACS (e)–(h) occupies an intermediate regime, while AMBER (a)–(d) spans a broad spectrum from ``cold'' to ``hot'' benchmarks. Cold AMBER benchmarks, such as TRPCage, saturate at lower power than compute-intensive cases like Cellulose or STMV. On modern GPUs (H100/H200), many workloads remain below the hardware power envelope even at default settings, including 5 of the 11 AMBER benchmarks.

\begin{figure}[t]
    \centering
    \subfloat[A100-AMBER]{%
        \includegraphics[width=0.47\linewidth]{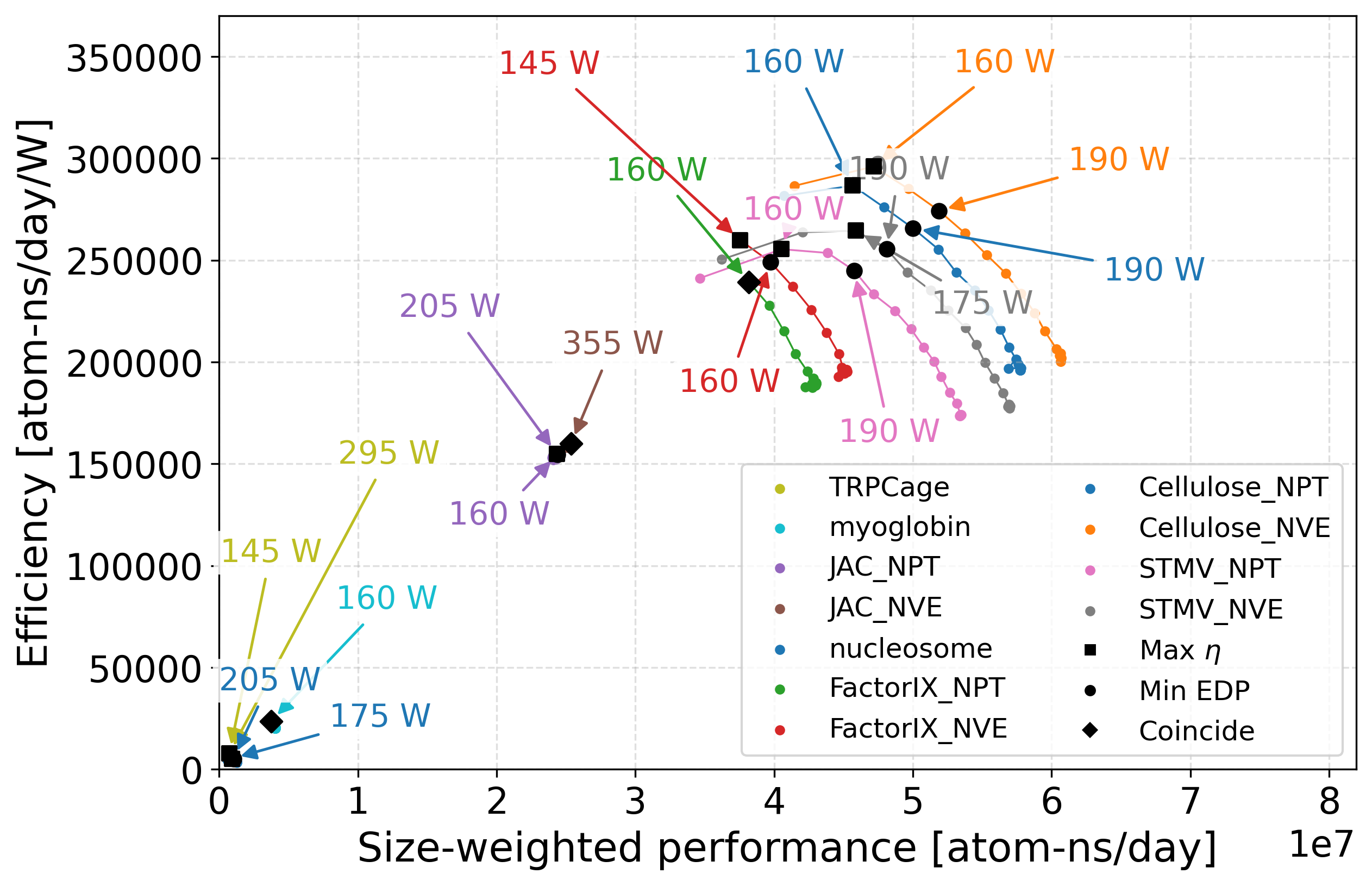}%
    }\quad
    \subfloat[H100-AMBER]{%
        \includegraphics[width=0.47\linewidth]{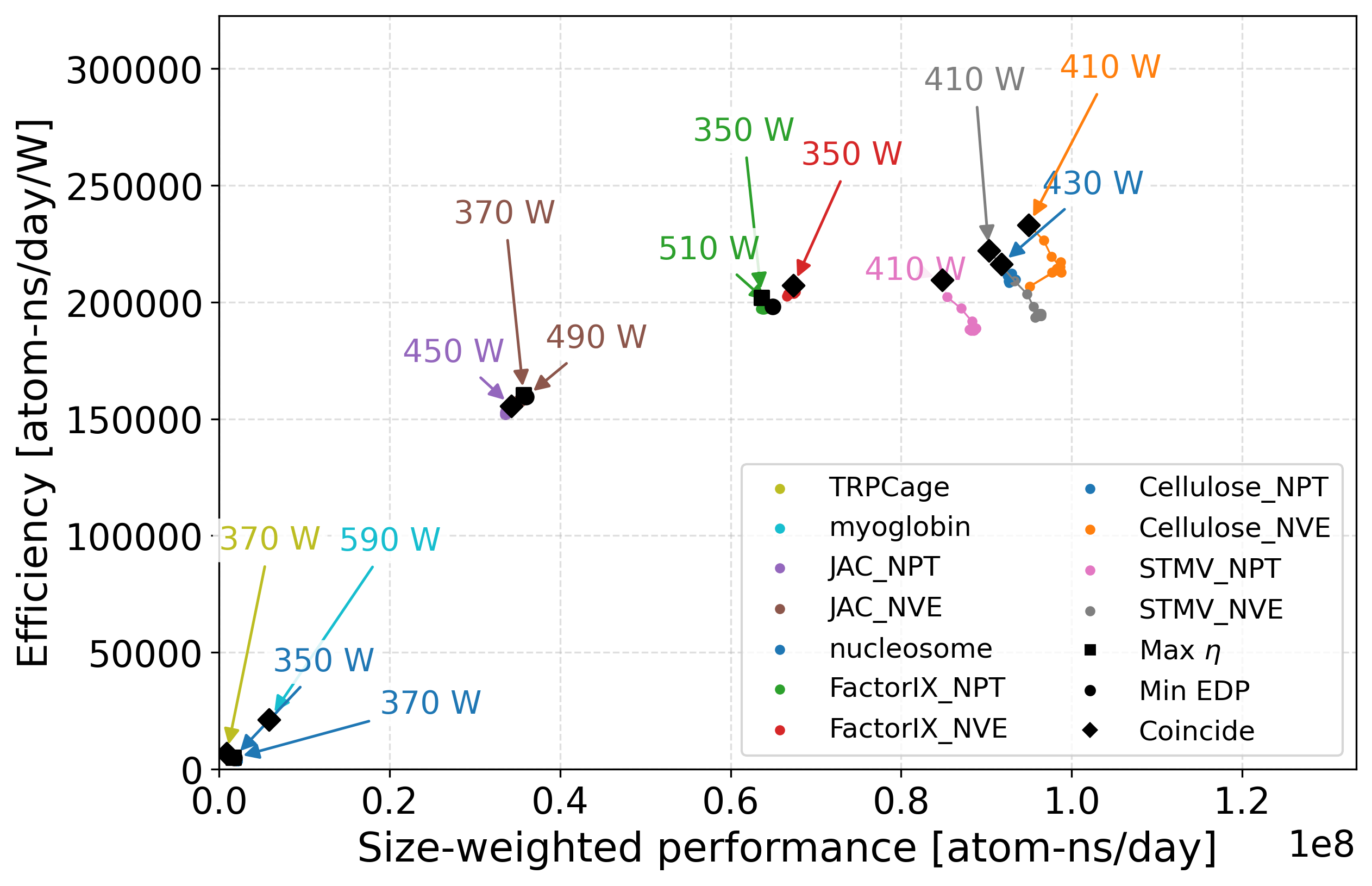}%
    }
\caption{Performance–energy efficiency Pareto space under power-cap settings for AMBER benchmarks on A100 and H100. Benchmarks are ordered from small to larger size, and optimal power caps for maximum energy efficiency and minimum EDP are labeled. Plots for the remaining workloads and GPUs are available at [\href{https://github.com/AyeshaAfzal91/GPU-Energy-Sweetspot/blob/main/PaperFigures/Fig7-All-GPUs-Workloads.pdf}{Link}].}
\label{fig:powercap}
\end{figure}
Figure~\ref{fig:powercap} demonstrates that power capping is generally less effective than frequency scaling for improving energy efficiency on modern GPUs and is strongly workload-dependent. The range of optimization through power caps is much smaller than achievable via frequency tuning. This difference is more pronounced on newer architectures: on older A100 GPUs (Fig.~\ref{fig:powercap}(a)), power capping AMBER behaves similarly to frequency tuning in terms of energy efficiency and performance (Fig.~\ref{fig:Zplot}(b)), with the minimum-energy configuration at $f^\ast_\eta = 990$\,MHz/ $W^\ast_\eta =160$\,W.
It show mixed AMBER behavior: larger benchmarks sometimes enter power-limited regions, while smaller benchmarks remain largely insensitive to cap reduction. This is most pronounced for compute-light workloads, which underutilize GPU resources and remain below the hardware power limit. Consequently, power caps offer limited optimization leverage, whereas frequency reduces dynamic power and improves efficiency across all regimes. On H100 (Fig.~\ref{fig:powercap}(a)), workloads rarely reach the hardware limit, so reducing the cap seldom triggers throttling or affects execution.
Unlike power capping, frequency tuning reshapes the entire power–performance trajectory, enabling workloads to traverse operating regimes. This explains why frequency mostly provides larger energy-efficiency gains at minimum-energy configurations. Overall, power-cap tuning primarily affects the high-performance points, whereas frequency scaling enables broader movement toward energy-efficient operating points.

\begin{figure}[t]
    \centering
    \subfloat[Power-frequency]{%
        \includegraphics[width=0.32\linewidth]{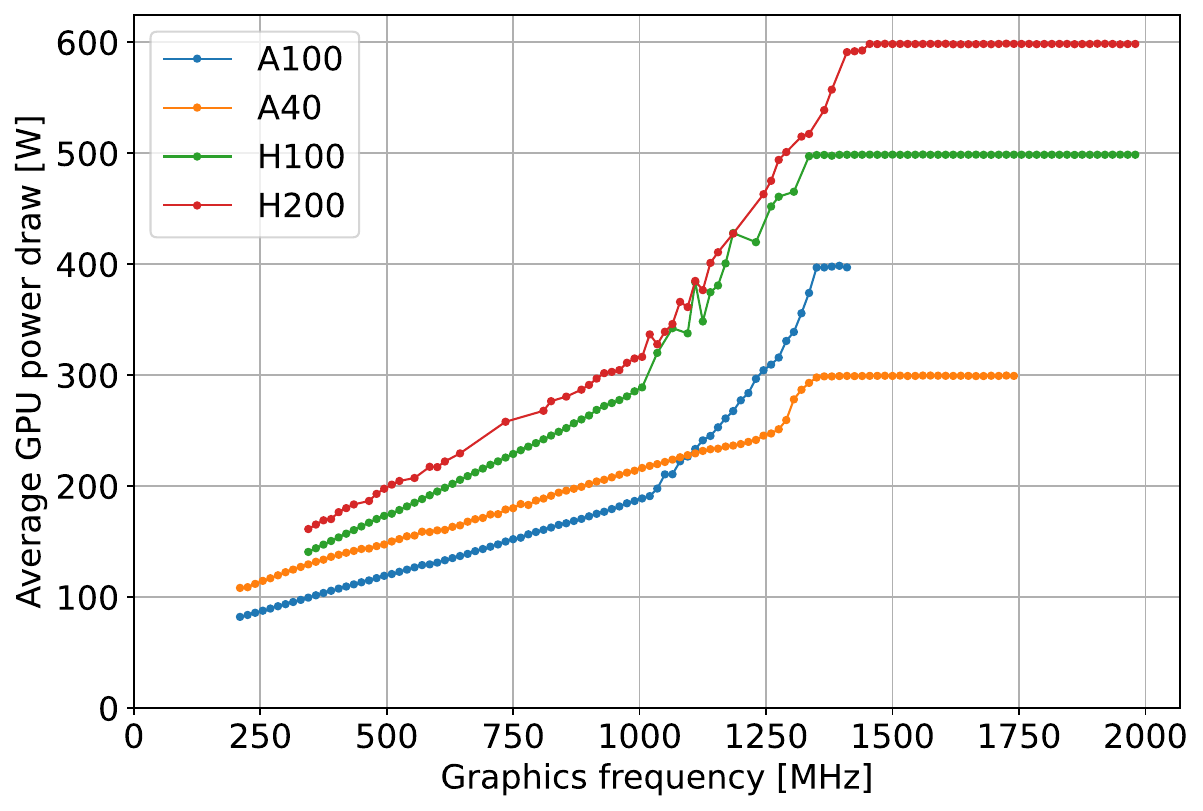}%
    }\quad
    \subfloat[Frequency-efficiency]{%
        \includegraphics[width=0.32\linewidth]{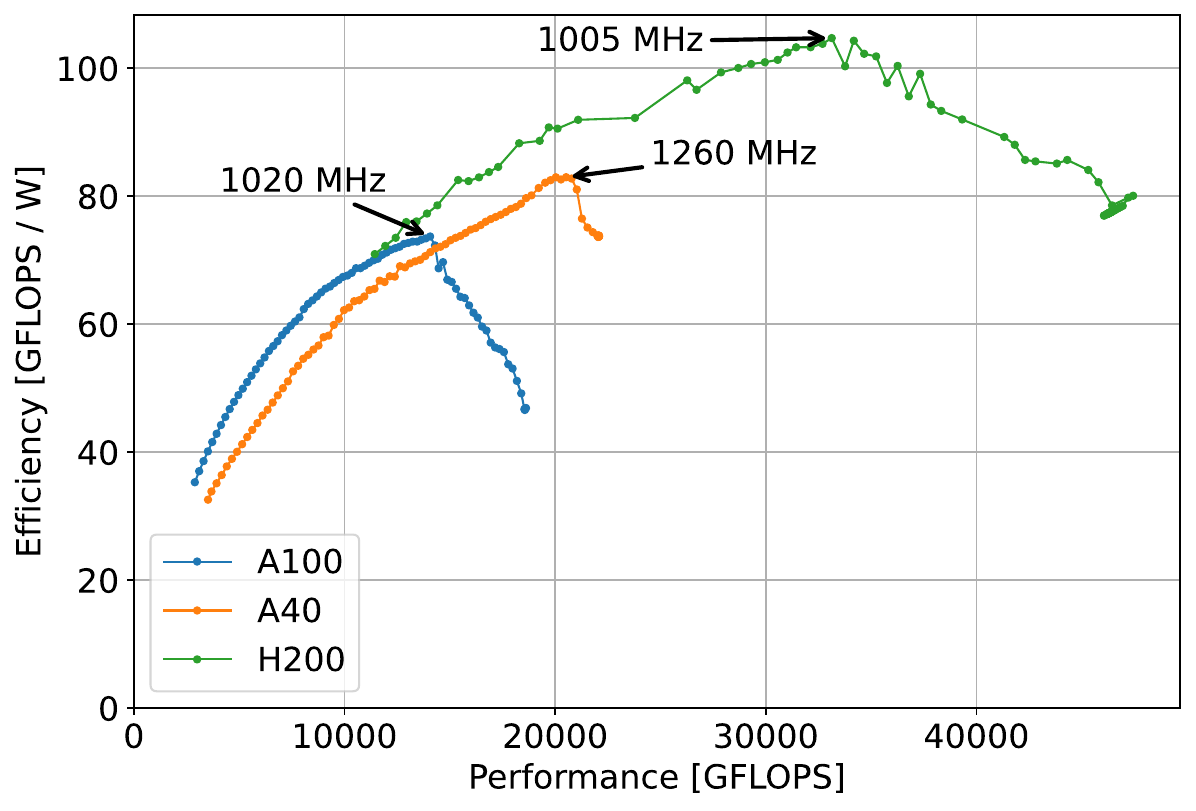}%
    }
    \subfloat[Powercap-efficiency]{%
        \includegraphics[width=0.32\linewidth]{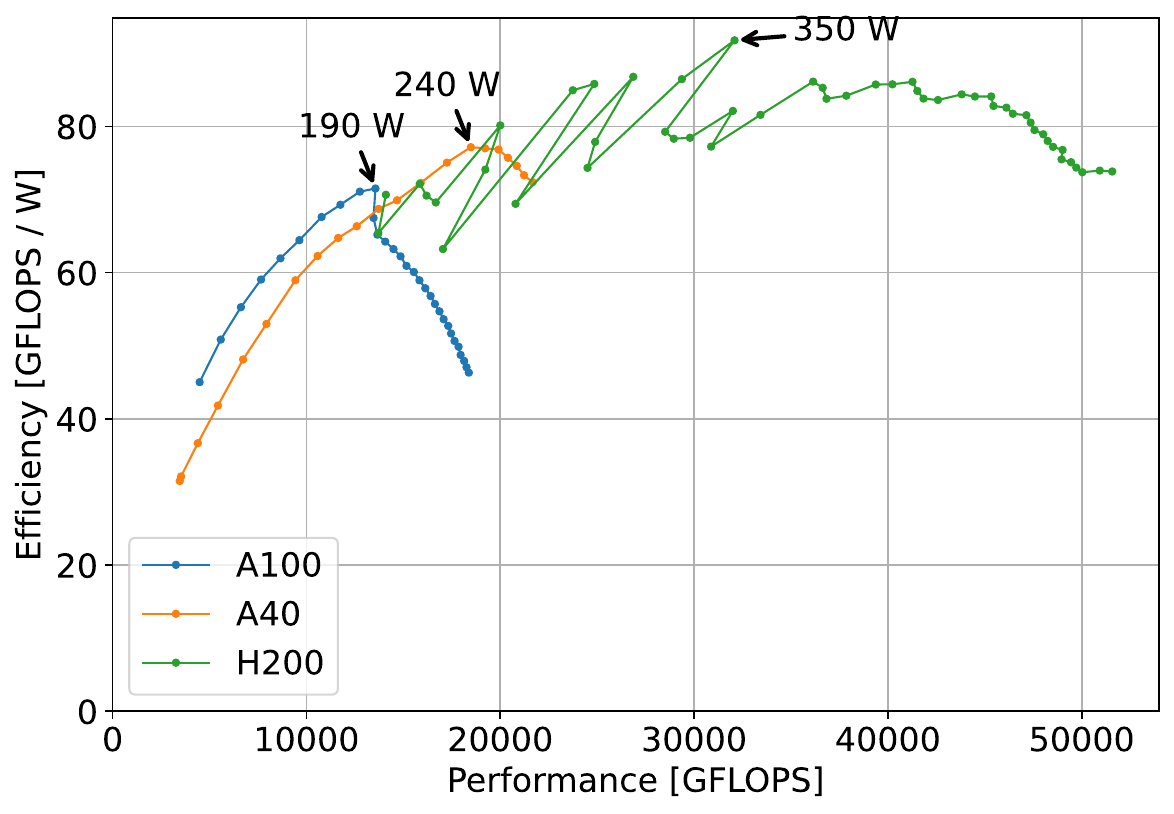}%
    } 
    \caption{(a) Piecewise power--frequency behavior and energy efficiency versus performance under (a) frequency scaling and (b) power capping for the FIRESTARTER GPU stress-test benchmark.}    \label{fig:firestarter}
\end{figure}
\textit{\textbf{Power capping vs. frequency scaling under stress}}\quad
On modern GPUs, power capping primarily affects thermally intensive (``hot'') workloads such as FIRESTARTER, while having moderate effects on AMBER hot benchmarks and negligible impact on ``cold'' (MD) workloads. 
Figure~\ref{fig:firestarter}(a) highlights the power-frequency behavior and saturation across architectures. Panel (b) demonstrates that frequency scaling consistently achieves the highest peak energy efficiency under high-stress conditions across all GPU generations (A40, A100, H200). 
Overall, frequency scaling yields the best performance-per-watt trade-off (e.g., 100\,GF/W vs.\ 80\,GF/W on H200; see Panel (b)-(c)). Consequently, H200 maintains performance above 48\,TFLOPS with stable efficiency under power-capped operation.  
For thermally intensive workloads, frequency scaling remains the most effective strategy: these workloads push the GPU toward its power envelope, invoking the cap mechanism to regulate frequency and voltage. In this regime, power capping yields only minor performance degradation and smaller energy-efficiency gains compared to frequency scaling.

\textit{\textbf{Impact of optimization on the Pareto frontier}}\quad
Figure~\ref{fig:tradeoff} evaluates whether optimization improves the energy-performance trade-off by comparing energy-efficiency gain $\Delta \eta$ against performance loss $\Delta P$. Points above the diagonal ($\Delta \eta > \Delta P$) indicate favorable trade-offs where efficiency improvements outweigh performance loss. 
Frequency scaling exposes a broader optimization space than power capping, enabling larger gains but with stronger workload- and architecture-dependent variability, as it reshapes the full power--performance trajectory. In contrast, power capping delivers smaller, more stable improvements by limiting only the upper power envelope.  
Older GPUs (A40/A100) generally achieve more favorable trade-offs across MD workloads, while newer GPUs (H100/H200) exhibit narrower efficient operating regions and greater sensitivity to nonlinear high-frequency effects and baseline/memory power. Larger workloads (yellow) consistently benefit from optimization, as fixed system costs are better amortized at high utilization.

\section{Conclusion and outlook}\label{sec:conclusion}
Across molecular dynamics and synthetic stress workloads, we show that energy efficiency is fundamentally a workload--hardware matching problem.
We propose a two-stage methodology: an analytic, piecewise GPU power model is first used to extract interpretable regime structure, which is then correlated with empirical measurements across real workloads and architectures. The model is fitted directly from application-level DVFS measurements, achieves low prediction error, and yields physically meaningful parameters that generalize across workloads.
In particular, baseline power $W_0$, nonlinear curvature $a_2$, and transition frequency $f_t$ define the energy landscape. The transition frequency $f_t$ marks the onset of inefficient operation, while $a_2$ controls the rate of efficiency degradation beyond this point. Across workloads, this structure consistently matches measurements: small workloads underutilize GPUs, medium workloads operate near optimal efficiency, and large workloads become bandwidth- or saturation-limited. While numerical values vary, the regime structure remains stable and transferable across architectures.
This shows that energy efficiency depends on aligning workload behavior with GPU regime boundaries. Frequency scaling is the dominant mechanism for navigating these regimes, reshaping the full power--performance frontier and enabling access to energy-optimal regions. In contrast, power capping provides limited benefit unless workloads approach sustained thermal or power limits on modern GPUs that often do not fully saturate hardware power envelopes.
Overall, energy-efficient HPC execution is a regime identification task defined by workload characteristics and a small set of interpretable hardware-dependent parameters.
Future work will extend this analysis to other accelerators such as AMD and Intel and emerging workloads such as large language models.

\subsection*{Acknowledgments}
The authors acknowledge support from the German Research Foundation (DFG), project number 545776403, FOR 5880: Holistic Energy and Performance Modeling for Sustainable Computing (Mod4Comp), as well as the EEC initiative within the national high-performance computing framework at German universities.
The authors gratefully acknowledge the HPC resources provided by the Erlangen National High Performance Computing Center (NHR@FAU) at FAU Erlangen-Nürnberg.
NHR funding is provided by the German Federal Ministry of Education and Research and the state governments participating on the basis of the resolutions of the GWK for the national high-performance computing at universities 
by federal and Bavarian state authorities. 
The NHR@FAU hardware is partially funded by the German Research Foundation (DFG), grant number 440719683.

\bibliographystyle{splncs04}
\bibliography{references-org}

\end{document}